\documentstyle[12pt]{article}
\begin{document}
\title{STANDARD SOLAR NEUTRINOS}
\author{Arnon Dar and Giora Shaviv\\
Department of Physics \\
and \\
Asher Space Research Institute \\
Technion-Israel Institute of Technology \\
 Haifa 32000, Israel \\
 arnon@technion.technion.ac.il \\
 gioras@physics.technion.ac.il}
\maketitle
\abstract
We describe in detail an improved standard solar model which
has been used to calculate the fluxes of standard solar neutrinos.
It includes premain sequence evolution, element diffusion,
partial ionization effects, and all the possible nuclear reactions
between the main elements. It uses updated values for the initial
solar element abundances, the solar age, the solar luminosity,
the nuclear reaction rates and the radiative opacities.
Neither nuclear equilibrium, nor complete ionization are assumed.
The calculated solar neutrino fluxes are compared with published
results from the four solar neutrino experiments. The
calculated $^8$B solar neutrino flux is consistent, within the
theoretical and experimental uncertainties, with the solar neutrino
observations at Homestake and Kamiokande. The observations
suggest that the $^7$Be solar neutrino flux is much smaller than that
predicted. However, conclusive evidence for the suppression of the
$^7$Be solar neutrino flux will require experiments like BOREXINO and
HELLAZ. If the $^7$Be solar neutrino flux is suppressed, it still
can be due either to standard physics and astrophysics
or neutrino properties beyond the standard electroweak model.
Only future  neutrino experiments, such as SNO, Superkamiokande,
BOREXINO and HELLAZ, will be able to show that the solar neutrino
problem is a consequence of neutrino properties beyond the standard
electroweak model.
 
\clearpage
\section{ Introduction}
\bigskip
The Sun is a typical main sequence star which
generates its energy by fusion of protons into Helium nuclei
through the pp and CNO nuclear reactions chains (Bethe 1939). These
reactions also produce neutrinos. These neutrinos have been detected
on Earth in four pioneering solar neutrino experiments, the
radiochemical Chlorine experiment at Homestake (Cleveland et al.
1995
and references therein), the electronic light water Cerenkov
experiment at Kamioka (Suzuki et al. 1995 and references therein)
and
the two radiochemical Gallium experiments, GALLEX at Gran Sasso
(Anselmann et al. 1995 and references therein) and SAGE at the
Baksan
(Abdurashitov et al. 1995 and references therein). They provide
the most direct evidence that the sun generates its energy via
fusion of Hydrogen into Helium. However, it has been claimed
(e.g., Bahcall 1995, Hata et al. 1995) that all four experiments
measured solar neutrino fluxes significantly smaller than those
predicted by standard solar models (SSM) (e.g., Bahcall and Ulrich
1988; Turck Chieze et al. 1988; Sackman et al. 1990, Bahcall and
Pinsonneault 1992 (BP92), Turck-Chieze and Lopes 1993 (TL93);
Castellani
et al. 1994; Kovetz \& Shaviv 1994; Christensen - Dalsgaard 1994;
Shi et al. 1994; Bahcall and Pinsonneault 1995 (BP95);
See, however, Dar and Shaviv 1994 (DS94); Shaviv 1995; Dzitko et al.
1995.
\smallskip
 
{\bf The Homestake chlorine experiment} has been collecting data
since 1970. Based on 25 years of observations
they have recently reported (Cleveland et al. 1995)
an average $^{37}$Ar production rate in $^{37}$Cl of
 
\begin{equation}
{\rm Rate~(Chlorine) =2.55\pm 0.17(stat)\pm 0.18(syst)}~SNU,
\end{equation}
(where $(1SNU=10^{-36}~s^{-1}$ captures per atom) by solar
neutrinos
above the $0.814~MeV$ threshold energy for the reaction
$\nu_e+^{37}$Cl$\rightarrow e^-+^{37}$Ar. It is
$32\pm 5\%$ of the $8.1^{+1.0}_{-1.2}~SNU$ predicted by the SSM of,
e.g., Bahcall and Pinsonneault 1992 (BP92).
\smallskip
 
{\bf Kamiokande II and III} observed electron recoils, with
energies first
above 9 MeV and later above 7 MeV, from elastic scattering of solar
neutrinos on electrons in water. Their 5.4-year data show a spectrum
consistent with $^8$B solar neutrino flux of (Suzuki 1995)
\begin{equation}
\phi_{\nu_\odot}=[2.9\pm 0.2(stat)\pm 0.3(syst)]
\times 10^6 cm^{-2}s^{-1}~,
\end{equation}
 which is $51\%\pm 9\%$ of that predicted by the SSM of BP92.
\smallskip
 
{\bf GALLEX}, the European Gallium experiment at the Gran Sasso
underground laboratory, measured a $^{71}$Ge production
rate by solar neutrinos in  $^{71}$Ga of (Anselmann et al. 1995a)
\begin{equation}
 {\rm Rate~(Gallium)=79\pm 10(stat)\pm 6(syst)}~SNU,
 \end{equation}
through the reaction $\nu_e+^{71}$Ga$\rightarrow e^++ ^{71}$Ge
whose
threshold energy is 233 KeV.
This rate is $60\%\pm 10\%$ of the
$131.5^{+7}_{-6}~SNU$ predicted by the SSM of BP92.
\smallskip
 
{\bf SAGE}, the Soviet-American Gallium Experiment in the Baksan
underground laboratory reported (Abdurashitov et al. 1995) an
average
$^{71}$Ge production rate by solar neutrinos in $^{71}$Ga of
\begin{equation}
 {\rm Rate~(Gallium)=74^{+13}_{-12}(stat)^{+5}_{-7}(sys)}~SNU
 \end{equation}
during 1990-1993, which is consistent with the rate reported by
GALLEX.
It is $56\%\pm 11\%$ of the $131.5^{+7}_{-6}~SNU$
predicted by the SSM of BP92.
\smallskip
 
These discrepancies between the observations and the predictions
have become known as
the solar neutrino problem(s). Three types of solutions to
these solar neutrino problems have been investigated:
 
(a) Terrestrial Solutions:
Perhaps the accuracy of the results of the
solar neutrino experiments has been
overestimated and unknown systematic errors
are largely responsible for the solar neutrino problem.
 
(b) Astrophysical Solutions:
Perhaps the standard solar models do not provide sufficiently
accurate description of the present day sun and/or the neutrino
producing
reactions in the sun.
 
(c) Particle Physics Solutions:
Perhaps non standard neutrino properties beyond the standard
model are responsible for the solar neutrino problem.
\smallskip
 
The chances that possibility (a) is responsible for the solar
neutrino problem have been greatly reduced by the GALLEX
Chromium
source experiment performed last year (Anselmann et al. 1995b).
This experiment is the first full demonstration of the reliability
of the radiochemical technique for the detection of solar neutrinos.
In particular, it excludes the possibility of any unidentified
important
sources of systematical errors, such as hot atom chemistry, in
the radiochemical experiments. However, we will argue that
terrestrial
solutions to the solar neutrino problem have not been ruled out. In
particular, the capture rate of solar neutrinos near threshold in
$^{37}$Cl and in $^{73}$Ga may be significantly smaller than
currently used for predicting SNUs (see chapter 7).
We also note that although Kamiokande, the electronic water
Cerenkov experiment, has definitely observed excess events pointing
towards the sun (Suzuki 1995), it has not been calibrated yet with a
known neutrino source. Such a direct calibration is highly desirable
since Kamiokande has the formidable task to single out solar
neutrino events with an expected rate of the order of
1/day in a fiducial volume of 680 tons of highly purified water
where the total trigger rate of the detector for electron
energy greater than 7 MeV is still 150,000/day.
\smallskip
 
It should also be noted that the joint observations of Homestake and
Kamiokande appear to be supported by the results of the Gallium
experiments
(Dar 1993). Namely, both the joint results from Homestake and
Kamiokande
and the results from GALLEX and SAGE seem to indicate that the
$^7$Be
solar neutrino
flux is much smaller than that predicted by the SSM (e.g., Dar 1993;
Hata
et al. 1994; Berezinsky 1994; Kwong and Rosen 1994; Parke 1995).
In fact,
the signals observed by GALLEX and SAGE can be predicted
accurately
from the observed solar luminosity and the solar neutrino
observations
of Homestake and Kamiokande (Dar 1993):
\smallskip
 
If the sun derives its energy from fusion of Hydrogen into Helium
and
if it is in a steady state where its nuclear energy production rate
equals its luminosity, then the total solar neutrino
flux at Earth is given by (e.g., Dar and Nussinov 1992)
\begin{equation}
\phi_{\nu_\odot}={2L_{\odot}\over Q-2\bar{E}_\nu}~{1\over 4\pi
D^{2}} \approx  6.51\times 10^{10}~cm^{-2} s^{-1}~,
\end{equation}
where $D\approx 1.496 \times 10^{13}~cm$ is the distance to the
sun,
$Q=26.733~MeV$ is the energy released when four protons fuse into
Helium,
and the average energy of solar neutrinos
has been approximated by  $\bar{E}_\nu(pp)
=\sum {E_{v,i}}\phi _{\nu ,i}\/ \sum {\phi _{\nu ,i}}
\approx
0.265~MeV$,
the average  energy of the $pp$ neutrinos which dominate the sum
rule.
If the pep and CNO contributions to the solar neutrino flux are
ignored then,
\begin{equation}
\phi_{\nu_\odot}\approx \phi_{\nu_\odot}(pp)
+\phi_{\nu_\odot}(^7Be)+\phi_{\nu_\odot}(^8B).
\end{equation}
Kamiokande reported
\begin{equation}
\phi_{\nu_\odot}(^8B)=(2.9\pm 0.4)\times
  10^6~cm^{-2}s^{-1}.
 \end{equation}
Thus, the flux of $^8$B neutrinos alone contributes
$<\sigma\phi>=3.2\pm 0.50~SNU$ to the $^{37}Ar$ production rate
in $^{37}$Cl, assuming a cross section of
$(1.11\pm 0.03)\times 10^{-42}cm^2$ for the capture of $^8$B
neutrinos
by $^{37}$Cl (Aufderheide et al. 1994).
The Chlorine experiment at Homestake which is
sensitive also to $^7$Be neutrinos observed
since 1970 an $^{37}$Ar production rate of only $2.55\pm
0.25~SNU$, and
$2.78\pm 0.35~SNU$ in runs 91-124
during 1986-1993, the running
period of Kamiokande II and III. For standard solar neutrinos
these results leave very little room for a $^7$Be
solar neutrino flux (the SSM estimates its expected contribution at
a level of $\sim 1~ SNU$). Therefore, we conclude (Dar 1993) that
$\phi_{\nu_\odot}(^7Be)\ll \phi_{\nu_{SSM}}(^7Be)$ and
$\phi_{\nu_\odot}\approx \phi_{\nu_\odot}(pp)$. Consequently,
the expected $^{71}$Ge production rate in $^{71}$Ga is given by
\begin{equation}
<\sigma \phi_{\nu_\odot}>_{Ga} \approx
<\sigma \phi_{\nu_\odot}(L_\odot)>_{Ga}+
<\sigma \phi_{\nu_\odot}(KAM)>_{Ga}
=83~SNU,
\end{equation}
where the first term on the rhs is the contribution of
the pp solar neutrino flux estimated from the solar luminosity and
the second term
is the contribution from the $^8$B solar neutrino flux measured by
Kamiokande. This predicted signal is in excellent agreement with
the $79\pm 12~SNU$ measured by GALLEX and the
$73\pm 19~SNU$ measured by SAGE, respectively.
\smallskip
 
Note, however, that in view of the experimental uncertainties,
the observed signal in the Chlorine experiment is not significantly
smaller than the expected minimal signal due to the solar neutrino
flux measured by Kamiokande. Thus, the solar neutrino experiments
do not provide conclusive evidence for new electroweak physics
as claimed, for instance, by Bahcall and Bethe (1991). Although
neutrino oscillations, and in particular the MSW effect (Mikheyev
and Smirnov 1986, Wolfenstein 1978) can solve the solar
neutrino problems, the inferred neutrino mixing parameters seem to
differ  substantially  from those implied by the atmospheric neutrino anomaly
(for a recent summary see e.g., Barish 1995)
or those implied
by the neutrino anomaly observed by the LSND collaboration
at LAMPF (Louis 1995; Athanassopolous et al. 1995; see, however
Hill 1995) . Moreover, in spite of extensive laboratory
searches and many precision tests no confirmed evidence from
accelerator experiments has been found for new physics beyond
the standard electroweak model (e.g., Langacker 1995
and references therein). Without such an
evidence it is quite important to improve the standard
solar model (which provides only an approximate description of the
complicated sun) and to continue the search for a
standard  physics solution to the solar neutrino problems.
\smallskip
 
In this paper we present an improved standard solar model and
its predictions for the solar neutrino fluxes.
The model includes premain sequence evolution, diffusion of all
elements,
partial ionization effects, and all the significant nuclear reactions
between the various elements which the sun is made of.
It uses updated values for the initial
solar element abundances, the solar age, the solar luminosity,
the nuclear reaction rates and the radiative opacities.
Neither nuclear equilibrium, nor complete ionization are assumed.
It employs a very fine zoning of the sun and accurate numerical
procedures to integrate the solar evolution equations from zero
age until the present day.
The calculated solar neutrino fluxes are compared with those
measured by the four solar neutrino experiments. The
calculated $^8$B solar neutrino flux is consistent, within the
theoretical and experimental uncertainties, with the solar neutrino
observations at Homestake and Kamiokande. However, the
observations
appear to
suggest that the $^7$Be solar neutrino flux is much smaller than that
predicted by our solar model. In spite of the facts that
the value of the $^7$Be solar neutrino flux is a robust prediction of
the current standard solar models and that
minimal extensions of the standard electroweak model, such as
neutrino
flavor mixing, can explain its suppression, as will be explained in
chapter 7, we do not consider the $^7$Be deficit
to be a compelling evidence for new electroweak physics,
as claimed by various authors (e.g., Bahcall and Bethe 1993;
Bludman et al. 1993; Castellani et al. 1994; Hata et al. 1994;
Berezinsky 1994; Kwong and Rosen 1994;
Bahcall 1994; Parke 1995; Hata and Langacker 1995). This is because
neutrino absorption cross sections near threshold in $^{37}$Cl
and in $^{73}$Ga may be significantly smaller than those calculated
by Bahcall (1989), because the
standard solar models are only approximate and simplified
descriptions of the real and complex sun, and because very little
is known, either experimentally or theoretically, on dense plasma
effects on nuclear reaction rates, decay rates, particle
and energy transport at solar conditions.
 
\smallskip
 
Our paper is organized as follows: In Section 2 we outline the
standard solar model. In Section 3 we describe the astrophysical
parameters which we use in this work. The main physics input is
described
in Section 4. Section 5 outlines our stellar evolution code. Our main
results are described in Section 6 and are compared there with
experimental results and the results of other solar model
calculations.
Possible standard
physics solutions to the solar neutrino problem are briefly outlined
in
section 7. The neutrino mixing parameters suggested by the MSW
solution are derived analytically in section 8. Final conclusions
are drawn in section 9.
 
\section{ The Standard Solar Model - An Overview}
\smallskip
The standard solar model (e.g. Bahcall 1989 and references therein)
is a physical description of the sun based on
{\bf the standard stellar evolution equations}, (e.g., Clayton 1968)
which are used to calculate its evolution from the
premain sequence Hayashi phase to its present state, using
{\bf the best available input physics} (initial conditions,
equations of state, nuclear cross sections, radiative opacities,
condensed matter effects).
The model assumes a complete spherical symmetry, no mass loss or
mass accretion, no angular momentum gain or loss, no differential
rotation and a zero magnetic field through the entire solar evolution .
Thus, the {\bf assumed initial conditions} are:
 
\begin{enumerate}
 
\item  {Fully convective, homogeneous, spherically symmetric
 protostar.}
 
\item {Initial mass of $M_\odot=1.99\times 10^{33}gm$.}
 
\item { No angular momentum, no differential rotation, no magnetic
field.}
 
\item {Initial chemical composition deduced from primitive
meteorites,
the solar photosphere, the solar wind, the local interstellar
gas and the photospheres of nearby stars.}
\end{enumerate}

The calculations are iterated, treating the unknown initial $^4$He
abundance and the mixing length in the convective zone (roughly the
size
of the pressure scale height) as adjustable parameters, until {\bf the
present day properties of the sun} are reproduced. These include:
 
\begin{enumerate}
 
\item {The solar luminosity $L_\odot=3.844\times 10^{33} erg\cdot
sec^{-1}.$}
 
\item { The solar radius $R_\odot=6.9599\times 10^{10}cm.$}
 
\item { The observed solar surface element abundances.}
 
\item { Internal structure consistent with helioseismology data
(optional).}
\end{enumerate}

The output of the calculations includes the present-day density
profile $\rho(r)$, temperature profile $T(r)$ and chemical
composition
profile $[X_i(r)]$ of the sun. They can be compared with information
extracted from helioseismology. They can also be used to calculate
the
expected fluxes of solar neutrinos. In particular,
according to the standard model, solar neutrinos are
produced mainly in the fusion of hydrogen into deuterium
($p+p\rightarrow D+e^++\nu_e$ and  $p+e+p\rightarrow D+\nu_e$),
in electron capture by $^7$Be and in $\beta$ decay of $^8$B,
$^{13}$N, $^{15}$O and $^{17}$F. Their production rates in the sun
are calculated using the standard electroweak theory and the
density, temperature and element abundances in the sun provided
by the standard solar model.
\smallskip
Our calculations were performed with an updated version of the
the solar evolution code of Kovetz and Shaviv.
We refer the reader to its detailed description by Kovetz and Shaviv
(1994) and focus here only on main points and important
improvements.
 
\section{  The Astrophysical Parameters}
\bigskip
\subsection{  Solar Luminosity}
\smallskip
The absolute luminosity of the sun changes with time due to
solar evolution. In addition to this secular solar variability
the solar luminosity changes with solar activity. The solar activity,
its period and erratic disappearance (last time during the Maunder
minimum) are not well understood. Precise
measurements of the solar ``constant'' during a large fraction of
an entire solar cycle have been carried out on a variety of
spacecrafts
and satellites during the last two solar cycles (21 and 22).
They include Nimbus 7 spacecraft (see, e.g., Hickey et al. 1982),
the Solar Maximum Mission Satellite (see, e.g.,
Wilson and Hudson 1988), The Earth Radiation Budget Satellite,
the NOAA9 and NOAA10 satellites (see e.g.,  Lee et al. 1991)
and The Upper Atmosphere Research Satellite (see e.g., Wilson 1993).
Using the following parametrization of the variation of the solar
constant ($SC$) during the solar cycle (Wilson and Hudson 1988),
\begin{equation}
 SC(t)=S\times(1+b(cos[2\pi(t-1980.82)/10.95]))~ W\cdot m^{-2},
\end{equation}
we found a weighted mean average $S=1367.2\pm 3.4 ~W\cdot m^{-
2}$
for the solar constant. It yields a mean solar luminosity of
$\L_\odot=4\pi D^2S=3.844\times 10^{33}~erg\cdot s^{-1}.$ This
value
is consistent with the value quoted by the Particle Data Group (1994)
and  used in BP95. It is smaller by 0.4\% than the value
$\L_\odot=3.86\times 10^{33}~erg\cdot s^{-1}$ used in BP92
and larger by 0.5\% than $L_\odot=3.826\times 10^{33}~erg\cdot s^{-
1}$
quoted before by the Particle Data Group (1992) and used by Dar and
Shaviv (1994).
 
\subsection{ Solar Age}
\smallskip
The solar age is best estimated from radioactive dating of meteoritic
condensation based on the observed $^{235}$U/$^{238}$U and
$^{207}$Pb/$^{206}$Pb ratios. Tilton (1988) estimated this age to be
$4.56\times 10^{9}y$. Guenther (1989) estimated it to be
$4.49\times 10^9y$
while recent analyses by Gopel et al. (1994) and by Wasserburg et al.
(1995) yielded $t_\odot=(4.566\pm 0.005)\times 10^9y$. This value
was used by us as the total solar age which includes both the
premain sequence evolution (some 30-40 million years) and the time
spent
by the sun on the main sequence.
 
\smallskip
\subsection{  Initial Chemical Composition}
\smallskip
The initial element abundances influence significantly the solar
evolution and the present density, chemical composition and
temperature
in the solar core, which determine the solar neutrino fluxes.
In particular, the calculated radiative opacities, which in turn
determine the
temperature gradient in the solar interior, are very sensitive
to the heavy elements abundances (the heavy elements are not
completely
ionized in the sun).
\smallskip
There are four major sources of information on the initial solar
abundances.
They are: the chemical composition of the most primitive class
of meteorites (type I carbonaceous chondrites), the solar
photospheric
abundances, the chemical composition of the solar wind and the
local interstellar medium element abundances.
\smallskip
\subsubsection{ The Heavy Metals Abundances}
\smallskip
Apart from the noble gases, only a few elements, namely, H, C, N and
O,
have escaped complete condensation in primitive early solar system
meteorites because they were able to form highly volatile
molecules or compounds (see, e.g., Sturenburg and Holweger 1990).
Thus, the initial solar abundances of all other elements are expected
to be approximately equal to those found in type I carbonaceous
chondrites
as a result of their complete condensation in the early solar system.
Since the chemical composition of the solar surface is believed to
have changed only slightly during the solar evolution (by nuclear
reactions during the Hayashi phase, by diffusion
and turbulent mixing in the convective layer during the main
sequence
evolution, and by cosmic ray interactions at the solar surface)
it has been expected that the photospheric abundances of these
elements
are approximately equal to those found in  CI chondrites.
Over the past decades there have been many initial disagreements
between the meteoritic and photospheric abundances.
In nearly all cases, when the atomic data were steadily improved
and
the more precise measurements were made, the photospheric values
approached the meteoritic values. The photospheric abundances are
now as a rule in very good agreement with the meteoritic values
(Grevesse and Noels 1993a). Hence, as meteoritic
values  represent the initial values and  are
known with much better accuracy (often better than 10\%) than
the photospheric ones, we have assumed that the initial solar
heavy metal abundances are given approximately by the meteoritic
(CI chondrites) values of Grevesse and Noels (1993a).
\vfill
\clearpage
\smallskip
\subsubsection{ The CNO Abundances}
\smallskip
The careful and comprehensive analysis of atomic and molecular
lines of C, N and O in the solar photosphere by Lambert (1978)
has served as the standard source of their solar abundance.
Improved atomic data, NLTE corrections
and study of infrared lines have slightly changed the standard values
of the photospheric CNO abundances to those given in Table I.
These values are very much different from those found in
primitive meteorites. In particular Carbon and Nitrogen are
underabundant in CI chondrites by about an order of magnitude
compared with their photospheric abundances (e.g., Sturenberg
and Howlweger 1990; Grevesse et al. 1990).
We have assumed in our calculations
that the initial CNO abundances are given
by the photospheric abundances properly corrected for diffusion
during the entire solar evolution.
\smallskip
\subsubsection{ The Abundances of $^4$He and $^3$He}
\smallskip
Because the sun is a G2 star its photospheric Helium abundance
can not be measured directly. The photospheric Helium abundance of
much
younger hotter
B stars in the solar neighborhood can
only be used for rough estimates of its initial solar abundance.
The initial $^4$He
mass fraction  in the solar nebula is known
only approximately, $0.24<Y<0.30$. The predictions of the
solar models are rather sensitive to the initial mass fraction of
$^4$He.
Consequently, the initial $^4$He solar abundance has been treated
in the standard solar models as an adjustable parameter.
The present day $^4$He surface mass fraction can be inferred
from helioseismology data which yield a surface
$^4$He mass fraction of $Y_s=0.242\pm 0.003$ (Hernandez and
Christensen-Dalsgaard 1994). However, their formal error is
highly misleading because of the great sensitivity of the result to the
model of the solar atmosphere, the equation of state there and
the atmospheric opacities. We estimate that at present
the $^4$He mass fraction at the solar surface is not known
from helioseismology better than $Y_s=0.242\pm 0.030$.
 
From measurement of the [$^3$He]/[$^4$He] ratio in meteorites and
in the solar wind and from the above rough estimate of the initial
abundance of $^4$He,
Geiss (1993) concluded that the presolar
abundance of $^3$He is, [$^3$He]/[H]=$(1.5\pm 0.3)\times 10^{-5}$
(by numbers), i.e., log([$^3$He]/[H])+12=$7.18\pm 0.08$.
 
\smallskip
\subsubsection{ The Deuterium Abundance}
\smallskip
Deuterium is easily destroyed already at relatively low
temperatures.
Consequently, all the primordial deuterium has been destroyed in
the Hayashi phase and its initial abundance cannot be estimated
from the solar photosphere. The
Deuterium abundance in the local
interstellar medium was recently  measured with the Hubble
Space
Telescope (Linsky et al. 1993) to be,
[D]/[H]=$1.65^{+0.07}_{-0.18}\times 10^{-5}$
(by numbers), i.e., log([D]/[H])+12=$7.22\pm 0.05$.
This value is consistent with the initial solar value,
[D]/[H]=$(2.6\pm 1.0)\times 10^{-5}$,
obtained by Geiss (1993) from the analysis of solar wind particles
captured in foils exposed on the moon and from studies of primitive
meteorites, which we used.
\smallskip
\subsubsection{  The Abundances of $^7$Li, $^9$Be and $^{11}$B}
\smallskip
The photospheric abundances of $^7$Li, $^9$Be and $^{11}$B are
smaller
by a factor of nearly 150, 3  and 10, respectively, than their
meteoritic abundances. The origin of such large differences is still
not clear. However, the initial solar (meteoritic)
abundances of Lithium, Berilium  and Boron are very
small and do not play any significant role in solar evolution.
On the other hand, their depletion can provide a clue to the history
of the convection zone and the solar evolution.
\bigskip
\section { The Physical Input}
\bigskip
 
\subsection{ Nuclear Reaction Rates}
\smallskip
The cross sections for most of the nuclear reactions that play
important role in the sun
fall steeply when the energy drops below the Coulomb
barrier. At solar energies they become too small to be measured
directly in laboratory experiments. Consequently, they are
either calculated theoretically or extrapolated
from laboratory measurements at much higher energies.
Because the results of different experiments often differ
considerably, and because different theoretical models used for
the extrapolation often yield different results,
the nuclear reaction rates used in different standard solar
models depend on personal judgment. Below
we discuss briefly some of the problematics of
obtaining reliable thermonuclear reaction rates for solar
conditions and we explain our specific choices.
These choices are also
summarized in Table II and compared there with the choices of
BP95.
For all other reactions we have chosen to use
the most detailed and complete compilation
of thermonuclear reaction rates published by Caughlan and Fowler
(1988).
\vfill
\clearpage
\subsubsection{ The Problematics}
\smallskip
To eliminate from the extrapolation the strong energy
dependence due to Coulomb barrier penetration,
 cross sections are normally parametrized as
\begin{equation}
\sigma_{ij}(E)= {S_{ij}(E)\over E} e^{-2\pi\eta}~,
\end{equation}
where $2\pi\eta=31.29Z_iZ_j\sqrt{\mu/E} $ is the Sommerfeld
parameter, $Z_i$ and $Z_j$ being the charge numbers of the
colliding nuclei, $\mu$ their reduced mass in atomic mass units
and E their center-of-mass energy in keV. The factor $S_{ij}(E)$ is
expected
to vary rather slowly with energy. It is  usually extracted from the
measured cross section at laboratory  energies and extrapolated
to the lower solar energies, using either a polynomial fit or an energy
dependence that follows from a specific model for the reaction.
\smallskip
 
The uncertainties in the nuclear reaction rates at solar conditions
are still large due to (1) uncertainties in the measured cross sections
at laboratory energies (2) uncertainties in the extrapolations
of the $S_{ij}(E)$ from laboratory energies down to solar energies,
(3) uncertainties in dense plasma effects (screening,
correlations and fluctuations) on reaction rates.
Unfortunately, only for a few simple reactions (
${\rm p+p\rightarrow D+e^++\nu_e}$,  ${\rm p+e+p\rightarrow
D+\nu_e}$
and ${\rm e^-+{^7Be}\rightarrow {^7Li}+\nu_e}$) the cross sections
can be
calculated accurately from theory. For all other direct reactions
neither the microscopic methods (for reviews see, e.g., Langanke
1991) such as the Resonating Group Method
(RGM) and  the Generator Coordinate Method (GCM), nor the potential
models such as the Optical Model (OM) and the Distorted Wave Born
Approximation (DWBA), can predict  accurate and reliable  low
energy
cross sections.
For instance,
the RGM and the GCM, which are currently considered to be the
best theoretical methods for calculating direct nuclear reactions,
predict (see, e.g. Descouvemont and Baye 1994, Johnson et al. 1992)
 $S_{17}(0)\approx 25-30~eV\cdot b$ for the reaction
${\rm p+{^7}Be\rightarrow{^8}B+\gamma}$. However a simple
inspection of
the results of these models reveals that they
reproduce poorly the magnitude of the measured cross
section, the position of the resonance, the width of the resonance,
the height of the resonance and the observed shape of the cross
section
as function of energy. To avoid these discrepancies
only the energy dependence of these models has been used by
Johnson et al. (1992)
to extrapolate the measured cross  sections to $E=0$, yielding
$S_{17}(0)\approx 22.4~eV\cdot b~.$
This value has been used in BP92 and BP95.
However, the procedure used by Johnson et al. (1992) is rather
an ad hoc one and it is
questionable in view of the facts that (a) their model does not
reproduce
accurately enough the measured energy dependence of the cross
section at
lab energies, and (b) they have extrapolated an ``average cross
section''
obtained by averaging cross sections which differ by many standard
deviations and have different energy shapes: The cross sections
measured
by Kavanagh (1960) and by Parker (1968) differ by more than
$3\sigma$
from the cross sections
measured later by Vaughn (1970) and by Filippone (1983)
in the same energy ranges (${\rm [Kavanagh]/
[Filippone]=1.34\pm 0.11}$, ${\rm [Parker]/[Vaughn]=1.42\pm
0.13}$;
see, e.g., Gai 1995).
\smallskip
\subsubsection{ The Choice of Extrapolation}
\smallskip
Dar and Shaviv (1994) have pointed out
that sub-Coulomb radiative captures
and transfer reactions take place mainly when the colliding
nuclei are far apart and not when their centers overlap. They
argued that most of the energy dependence
of the astrophysical $S_{ij}(0)$  factors is because
Eq. 10 ignores this fact.
The Coulomb barrier penetration factor
$exp(-2\pi\eta)/\sqrt{E}$ in Eq. 10  actually represents
\begin{equation}
\vert\psi_c(0)\vert^2\approx {2\pi\eta\over e^{2\pi\eta}-1},
\end{equation}
the absolute
value squared of the Coulomb wave function at the origin
(normalized
asymptotically to a plane wave of unit amplitude), while
the physical cross section is proportional to the
square of the Coulomb wave function at the effective distance R
where the reaction takes place, i.e., to the Coulomb barrier
penetration factor at a relative distance R. They concluded
that the
optical models which describe well the shapes of the bound state
and relative motion wave functions outside the nuclear potential
are the most reliable for extrapolating
the laboratory cross sections to solar energies. Alternatively,
they proposed that after extracting the energy dependence of
the Coulomb barrier penetration factor to the effective distance R
where
the reaction takes place (and other trivial energy dependencies)
a simple polynomial fit  can be used to extrapolate the lab cross
section to solar energies.
 
\smallskip
Let us show that the reactions take place mainly when the nuclei
are far apart.
Consider for instance radiative captures. The transition amplitude
is proportional to an overlap integral $I=\int u^*(r)r\psi_c(r) dr$
where $u$ and $\psi_c$ are the radial parts of the bound state wave
function and of the Coulomb distorted wave function of the initial
relative motion, respectively. The bound
state wave function outside the nuclear radius falls exponentially
and has the asymptotic form, $u(r)\sim e^{-\beta r}$
where $\beta =\sqrt{2\mu E_b}/\hbar) $ with $E_b$ being the
nuclear
binding energy of the colliding nuclei in the final nucleus.
The incident Coulomb wave function decreases inside the
Coulomb barrier (in the WKB approximation) like
\begin{equation}
\psi_c(r)\sim [V_c(r)-E]^{-1/4}e^{-{\sqrt{2\mu}\over
\hbar}\int^{R_0}_r [V_c(r)-E]^{1/2}dr}~,
\end{equation}
where $V_c(r)$ is the effective Coulomb potential (Coulomb plus
centrifugal) and $R_0$ is the classical turning point.
At keV energies, E is much smaller than $V_c(r)$
near the nucleus and
\begin{equation}
\psi_c(r)\sim r^{+1/4}e^{-2\eta[\pi/2-arcsin(E/E_c)^{1/2}
-(E/E_c)^{1/2}(1-E/E_c)^{1/2}]}\sim r^{1/4}e^{\gamma \sqrt{r}},
\end{equation}
where $E_c=Z_iZ_je^2/r$ and $\gamma=\sqrt{8Z_iZ_je^2\mu
}/\hbar$.
Consequently, the contribution to the radial overlap integral
comes mainly from the vicinity of $r=R$, where
\begin{equation}
R\approx {\gamma^2\over(4\beta)^2}\left(1+\sqrt{1+{
20\beta
\over \gamma^2}}\right)^2\approx {\gamma^2\over4\beta^2 }=
{Z_iZ_je^2\over E_b}.
\end{equation}
As an example, for the reaction
${\rm ^3He+^4He\rightarrow ^7Be+\gamma}$ one finds
$R\approx 10~fm$, while for the
reaction
${\rm p+^7Be\rightarrow ^8B+\gamma}$ one finds
$R\approx 70~fm$  in good
agreement
with the Optical Model and DWBA calculations.
Eq. 14 is valid only if $R_N\leq R \leq R_0$, where $R_N$ is the
nuclear radius of the i+j bound state and $R_0=Z_iZ_j/E$ is the
classical
turning point.  If $R<R_N$ then
$R\approx R_N$. Similarly, if $R\geq R_0$ then
$R\approx R_0$.
Thus, instead of being proportional to the Coulomb barrier
penetration
factor at the origin, the radiative capture cross
section is proportional to
$k_{\gamma}^3\vert u(R)R\psi_c(R)\vert^2$.
One can divide out this
energy dependence and  use a polynomial fit to extrapolate the
reduced cross section to solar energies (Dar and Shaviv 1994)
or use instead the Optical Model or DWBA to extrapolate the total
cross section to solar energies. The resulting astrophysical $S_{i,J}$
factors are very similar. They are summarized here only for the
main reactions.
 
\subsubsection{ $S_{17}(0)$}
\smallskip
Extrapolations of the cross sections measured by Vaughn (1970) and
by Filippone(1983), using either simple potential models or the
above
very general properties of sub-Coulomb
cross section, gave $S_{17}(0)\approx 17~eV\cdot b$ (Barker and
Spear
1986; Dar and Shaviv 1994; Kim et al. 1994).
This value seems to be supported also by other types of experiments:
Analysis of recent measurements of the virtual reaction
$\gamma_v+^8$B$\rightarrow p+^7$Be through Coulomb dissociation
of
$^8$B in the Coulomb field of Lead (Motobayashi et al. 1994) gave
$S_{17}(0)\approx 17~eV\cdot b$. A similar value,
$S_{17}(0)\approx 17.6~eV\cdot b$, was estimated from the virtual
reaction ${\rm p_v+^7Be\rightarrow ^8B}$ measured
through the proton transfer reaction
$^3$He+$^7$Be$\rightarrow$D+$^8$B (Xu et al. 1994).
Consequently, we a have adopted the value $S_{17}=17~eV\cdot b$
in
our standard solar model calculations. The value
$S_{17}=22.4~eV\cdot b$ was used in BP92 and BP95.
\smallskip
\subsubsection{ $S_{34}(0)$}
\smallskip
The value
$S_{34}(0)= 0.51\pm 0.02~keV\cdot b $ was obtained from
measurements of the prompt $\gamma$-ray emitted in the reaction
${\rm ^3He+^4He}\rightarrow ^7Be+\gamma$, while measurements
of the induced $^7$Be activity led to a weighted average
$S_{34}(0)= 0.58\pm 0.02~keV\cdot b $, which are different by
3.5 standard deviations. The origin of this discrepancy is still
not known (Hilgemeier et al. 1988). Normalization to known
cross sections favor the lower value.
 
Using the measured energy dependence of
the cross section for ${\rm ^3He+^4He\rightarrow ^7Be+\gamma}$ by
Krawinkel et al. (1982) for extrapolating the cross sections measured
by prompt gamma ray emission
(Parker and Kavanagh 1963;
Osborne et al. 1982; Krawinkel et al. 1982 (multiplied by 1.4);
 Hilgemeier et al. 1988) to zero energy, we obtained
$S_{34}(0)= 0.45~keV\cdot b $. The value
$S_{34}(0)= 0.533~keV\cdot b $ was used in BP92 and
$S_{34}(0)= 0.524~keV\cdot b $ was used in BP95.
\vfill
\clearpage
\subsubsection{ $S_{33}(0)$}
\smallskip
 
A similar analysis of the low energy data of Greife et al. (1994),
Krauss et al. (1987) and Dawarakanath and Winkler (1971)
on the reaction ${\rm ^3He+^3He\rightarrow ^4He+2p}$
gave $S_{33}(0)=5.6~MeV\cdot b$. Essentially the same value,
$S_{33}(0)=5.57~MeV\cdot b$,
was obtained by Krauss et al. (1987)  and by Greife et al. (1944)
by applying a polynomial fit to their data.
The value $S_{33}(0)=5.0~MeV\cdot b$ was used in BP92 and
$S_{33}(0)=4.99~MeV\cdot b$ was used in BP95.
\smallskip
\subsubsection{ $S_{11}(0)$}
\smallskip
The cross section for the reaction
${\rm p+p\rightarrow D+e^++\nu_e}$ is too small
to be measured directly in the laboratory. The measured cross
sections
for the weak isospin
related reactions
$\bar{\nu}_e+$D$\rightarrow e^+$+n+n,
$\bar{\nu}_e+$D$\rightarrow\bar{
\nu}_e$+p+n, and $\gamma+$D$\rightarrow$n+p
were used to obtain the relevant nuclear matrix element needed
for calculating the cross section for p+p$\rightarrow$D+$e^++\nu_e$.
This procedure yielded  a best value
$S_{11}(0)\approx 4.07\times 10^{-22}~keV\cdot b$
which is consistent with the value used by Caughlan and
Fowler (1988). It is 4.2\% larger than the value
$S_{11}(0)\approx 3.896\times 10^{-22}~keV\cdot b$ calculated
recently
by Kamionkowski \& Bahcall (1994) and used in BP95.
 
\smallskip
Table II compares the astrophysical
$S_{ij}(0)$ factors used in this work (and also in DS94) and in BP95.
\smallskip
\subsubsection{ Screening Enhancement of Reaction Rates}
\smallskip
Screening of target nuclei by their electrons is known to enhance
significantly laboratory nuclear cross sections at very low
energies (e.g., Engstler et al. 1988)
although a complete theoretical understanding of the effect
is still lacking (see, e.g.,
Shoppa et al. 1993; Rolfs 1994 and references therein).
Screening corrections to the nuclear reaction rates in the SSM are
usually represented by an enhancement factor
\begin{equation}
 F\approx e^{\Delta U/kT}\approx e^{Z_iZ_je^2/R_DkT},
\end{equation}
where $\Delta U$ is the gain in electronic
potential energy when an incident ion of charge $Z_je$ penetrates
the charged  cloud around an ion of charge $Z_ie$, T is the plasma
temperature and $R_D$ is the Debye length,
$R_D\equiv (kT/ 4\pi e^2\Sigma Z^2\bar n_{_Z})^{1/2}. $
This screening enhancement of thermonuclear reaction rates
near the center of the sun is quite
considerable, being 5\%, for the pp and pep reactions, 20\% for
the ${\rm ^3He^3He}$, ${\rm ^3He^4He}$ and ${\rm p^7Be}$ reactions,
and 30\%, 35\% and 40\% for the p capture by the C, N and O
isotopes,
respectively.
 
The change in the Coulomb barrier (or equivalently the gain in
energy)
due to screening is estimated from the Debye-H${\ddot {\rm u}}$ckel (DH)
approximation
for the screened potential around {\it a static} ion in an electrically
neutral plasma ($\Sigma Z\bar n_{_Z}=0$,  $\bar n_{_Z}$ being
the number density of particles of charge $Ze$ with $Z=-1$ for electrons):
\begin{equation}
\Phi_i={eZ_i\over r}e^{-r/R_D}\approx {eZ_i\over r}-
   {eZ_i\over R_D} ~~{\rm for}~r\ll R_D,
\end{equation}
which is an approximate
solution to Poisson's equation
\begin{equation}
 \nabla^2\Phi_i=4\pi e\Sigma Z n_{_Z}=
4\pi e\Sigma Z\bar n_{_Z}e^{-eZ\Phi_i/kT}\approx R_D^2\Phi_i,
\end{equation}
near a static ion of charge $Z_i$. This equation and solution, however, are
valid
only far from the nucleus where $eZ\Phi_i\ll kT $, or else the expansion
is not valid.
In the core of the sun where $kT~(\sim 1~keV)$
is much smaller than the Coulomb barriers ($\sim 1~ MeV$) between
the reacting nuclei, most of the contribution to the nuclear
reaction rates comes from collisions with c.m. energies
$E\gg kT.$ At the classical turning point $eZ_j\Phi_i=E.$
Consequently, inside the barrier, $eZ_j\Phi_i\gg kT$,
and the naive use of the DH approximate potential
for calculating the barrier penetration factor (e.g. Clayton 1968)
is unjustified for projectiles with wavelengths much shorter than
the classical distance of closest approach.
The DH solution also assumes that there
are many ions and many electrons within a Debye sphere.
Also these conditions are not quite satisfied in the core
of the Sun, where the inter-ion distance
is about the same as the Debye length.
It was found that when the conditions for the
validity of the DH approximation are violated in
laboratory plasmas, calculations based on the DH approximation
fail dramatically (Goldsmith et al. 1984)
in reproducing various observations, unless
the wavelength of the particle under consideration is much larger
than the Debye length. The later is the case for the electron capture
reaction $^7$Be($e^-,\nu_e)^7$Li, where it was found by Johnson et
al.
(1992) that the DH screening potential changes
$\vert\psi_e(0)\vert ^2$ by only a few percent.
However, for the thermonuclear reactions in the sun the wave
lengths
of the reacting ions are very small in comparison with their classical
distances of closest approach and the Debye length.
 
Moreover, an ion approaching from infinity a target nucleus
gains from the Debye cloud a potential
energy $\Delta U\approx e^2Z_iZ_j/ R_D$
when it penetrates it. But, there is no gain in potential
energy if either its
initial position is already inside the Debye cloud
(where the cloud potential is constant $V\approx eZ_i/ R_D$)
or if the ion leaves and enters similar potential wells.
 
Near the center of the Sun, where
$\rho_c\approx 156~g~cm^{-3},$ $T_c \approx 1.57\times 10^7K$,
$X_c(H)\approx 34\%$, and $X_c(He)\approx 64\%$,
the average inter-ionic spacing is $n_i^{-1/3}\sim 2.8\times 10^{-
9}~cm,$
similar to the Debye length, $R_D\approx 2.3\times 10^{-9}~cm.$
On the other hand the gain in potential energy is increased
if both ions retain their Debye clouds during their approach.
However, the most effective ion energies are 5-20 kT
and their velocities are not much smaller than the average
thermal electron velocities. There may not be sufficient time for
the plasma to rearrange itself around the fast moving ions
and screen them effectively (the typical rearrangement
time, $2\pi/\omega_p$, where $\omega_p= (4\pi n_e^2/m_e)^{1/2}$,
is longer than the time it takes a fast ion
to cross the Debye length). All these effects may modify substantially
the screening enhancement of the fusion reaction rates near the
center
of the Sun as calculated from the static DH screening potential.
Actually,
the situation is too complicated for reliable analytic estimates.
Reliable estimates, however, may be based on numerical N body
simulation of the classical trajectories of electrons and ions in
a dense plasma which can be used to evaluate the various
effects of correlations and fluctuations on the thermonuclear
reaction rates in dense plasmas (Shaviv and Shaviv 1995).
\smallskip
To test the sensitivity of the standard solar models to the screening
corrections we have carried out the calculations with and without
the standard screening enhancement factors of all the thermonuclear
reaction rates. We found (Dar and Shaviv 1994) that
removing/including the screening enhancement factors for
{\bf all} nuclear reaction rates
had only a small effect on the calculated solar neutrino
fluxes, due to accidental cancellations.
Screening enhancement may, however, play important role in other
stars.
Moreover, changes in the screening factors which change their ratios
for the different reactions may cause significant changes in the
calculated solar neutrino fluxes.
\smallskip
\subsubsection{ Radiative Opacities}
\smallskip
The radiative opacities depend on the local chemical composition,
density and temperature in the sun. We have used radiative opacity
tables
computed by the OPAL group at Lawrence Livermore National
Laboratory
that were kindly provided to us by
F.J. Rogers. They are updated version of the OPAL tables of
Rogers and Iglesias (1992) for the most recent determination
by Grevesse and Noels (1993a)
of the heavy element composition of the sun from
the meteoritic and photospheric data.
As for the low temperature opacities, we used the Alexander, Rympa
\&
Johnson (1983) low temperature opacities which include the effect of
molecules. Note that the original table contains a numerical error.
This
error was corrected by interpolation from adjacent values. Since the
opacities are differentiated for the numerical scheme a spline
interpolation is used to guarantee regular derivatives and opacities.
Note that the nuclear reactions and the element
diffusion change the heavy element abundance throughout
the sun as well as its internal breakdown to the various elements.
The breakdown of the OPAL opacities to the
individual contributions of the various elements
as function of local density, temperature and composition is
not available yet for general use. However, the calculations
of Kovetz and Shaviv (1994) show that the diffusion of
the heavy elements is approximately the same for all the heavy
metals including the partial ionization of each element
(see e.g., Table I of Kovetz \& Shaviv 1994). Thus, the error introduced
in the radiative opacity by the neglect of the change in the heavy
element breakdown is expected to be negligible compared to other
inaccuracies in the opacities.
Besides diffusion the only important
process that changes the relative abundances of the heavy elements
is the conversion of $^{12}$C to $^{14}$N in the core of the sun.
However, since the radiative opacities of C and N are not very
different when they are completely ionized, we do not expect  this
conversion to affect significantly the total radiative opacity.
Consequently, we calculated the opacities for the values
of $\rho$, T, X and Z in each shell by
interpolating the OPAL opacities, assuming the same
breakdown of Z to the heavy elements. We plan to extend the present
calculations to include the effects of the variations of the heavy
element breakdown on the radiative opacity.
\smallskip
\subsubsection{ Equation of State and Partial Ionization}
\smallskip
Under the conditions prevailing in the core of the sun, the electrons
are partially degenerate. Their contributions to the various
thermodynamic
quantities (pressure, entropy, adiabatic exponents) is calculated
in the Kovetz Shaviv code by exact evaluation of the Fermi-Dirac
integrals. The Coulomb interactions in the solar plasma have been
included in the non ideal part of the free energy.
The value of the Coulomb parameter in the core of the sun is about
0.1-0.2.
Kovetz \& Shaviv (1994) and Dar \& Shaviv (1994)
 used a fit by Iben et al. (1992) to the
results
of Slattery, Doolen \& De Witt (1980) for the internal energy of the
gas
which were obtained for the Coulomb parameter $\Gamma
\le 1$. These results were calculated for a One Component Plasma
(OCP)
and ignore the contribution of the electrons. Recent results by Shaviv
and Shaviv (1996) indicate that the assumption of OCP is not so good
for
such low values of $\Gamma$ and hence, we calculated a model with
the
simple Debye-H${\ddot {\rm u}}$ckel model.
The local ionization state of each
element is calculated by solving the local Saha equations.
Partial ionization effects are included in the equation of state and
in the diffusion coefficients (C,N, O or Mg are once to twice ionized
at the surface and completely ionized in the core. Fe is doubly
ionized at the surface but only 16-20 times ionized in the core).
Partial ionization has an important effect on the equation of state
in the outer part of the sun even if the relevant region is convective
since the partial ionization affects the entropy of the gas.
 
\bigskip
\section{ The Solar Evolution Code}
\bigskip
Modeling the solar evolution requires integration of the standard
stellar
evolution equations over very long times. Truncations and
inaccuracies
in the time integration may add to large errors if not properly
handled.
In particular, the numerical integration must ensure energy
conservation.
Namely, the integral of the total solar luminosity in
photons and neutrinos over a long time
must be equal to the change in the total solar
binding energy (nuclear, gravitational, thermal and Coulomb) during
that time. A carefully constructed advanced stellar evolution code
was described by Kovetz and Shaviv (1994).
We have used this code to calculate the evolution of a 1
${M}_{\odot}$
protostar, fully convective with a uniform
composition, initial solar element abundances
and a luminosity around $12L_{\odot}$. The code follows the
Hayashi track, the settling to the main sequence (ZAMS)
and the main sequence evolution.
It includes hydrodynamic, gravitational and nuclear
energy release, radiative energy transport, convection, and diffusion
of all elements more abundant than a prescribed abundance.
 Neither nuclear equilibrium, nor complete ionization
are imposed. The effects of partial ionization are included in
the equation of state, the diffusion coefficients and the
radiative opacities of OPAL and in the calculation of the
solar atmosphere. For the convenience of the reader we highlight
here
some important details of the stellar evolution code.
For more details the reader is referred to its description
in the original paper of Kovetz and Shaviv (1994).
\smallskip
\vfill
\clearpage
\subsection{ Nuclear Evolution}
\smallskip
The nuclear evolution equations for the local number fractions
$y_i=X_i/A_i$  have the general form
\begin{equation}
{dy_i\over dt}=\Sigma_kS_{i,k}({\bf y})y_k
\end{equation}
where $S({\bf y})$ is a matrix with positive (creation) and
negative (destruction) elements.
Most of these elements are linear in the $y$'s, but some
(corresponding to beta decays) are constants. Thus, the right hand
side of Eq. 18 is quadratic in the $y$'s.
The basic nuclear
network contains all the isotopes and nuclear reactions which appear
in
Caughlan and Fowler (1988). The nuclear reactions rates have been
updated
with the reaction rates of Wiescher et al. (1989), Kubono et al.
(1989) and Gorres et al. (1989). The other changes (updating the
normalization of the ${\rm ^3He^3He,~^3He^4He, p^7Be}$ reactions)
were described in section 4. The equation is solved by conversion
into a finite difference equation. Stability and the requirement that
the solution approaches the solution for nuclear equilibrium as time
tends to infinity require that the time differencing be implicit.
Thus, Eq. 18 is replaced by the fully implicit finite difference
system,
\begin{equation}
 {\bf y}_{t+\delta t}={\bf y}_t+\delta t S({\bf y}_{t+\delta t})
{\bf y}_{t+\delta t},
\end{equation}
which is correct to $O(\delta t)$, numerically stable and delivers
the correct values $y_i(t+\delta t)$ for those isotopes that
change very rapidly and are thus in nuclear equilibrium.
No special provision is required for handling isotopes that are in or
tend towards nuclear equilibrium. The non linear equations 19
are solved by iterations in which large matrices are inverted
(which consume a lot of CPU time). About three iterations
were needed to obtain an accuracy of $1:10^{-10}$ needed in order
to follow the concentration of species like $^7$Li and $^8$B.
Our nuclear network contains about 80 isotopes, of which only
the following 18 isotopes were found to play important role
in the nuclear evolution:
${\rm ^1H,^2D,^3He,^4He,^7Li,^9Be,^8B,^{12}C,^{13}C,^{13}N,^{14}N,
^{15}N,^{15}O,^{16}O,^{17}O,^{17}F,^{18}F,^{20}Ne,}$
 ${\rm^{23}Na.}$
The nuclear network searches automatically for the most important
contributing reactions at each phase. In the particular case of the sun
the network  was reduced to these 18 isotopes. Yet, reactions like
$^{3}He^2D$ were included by the program.
(The other elements, in particular Ne, Na, Mg, Si, S and Fe,
which are present in the sun
in significant quantities that affect the ``heavy element'' abundance Z
but do not participate in the nuclear reactions,
were included in the calculations of the opacities,
the equation of state, the thermodynamic quantities
and the diffusion of elements).
\smallskip
The nuclear energy generation rate $Q_N$, which is the time
derivative
of the total nuclear binding energy is calculated from:
\begin{equation}
 Q_N=-{\bf E}_b\dot {\bf S}({\bf
y}_{t+\delta t}){\bf y}_{t+\delta t},
\end{equation}
where ${\bf E}_b$ is the vector of nuclear binding energies.
This form ensures that the integral over time of the nuclear
energy released in the sun is equal to the change in the solar
nuclear binding energy during that time.
\smallskip
\subsection{Space and Time Steps}
\smallskip
The sun was divided into 550 mass shells, distributed in such away
to yield sufficiently fine distribution in the core, near the bottom
of the outer convective zone (where $\Delta m$ was
$1\times 10^{-4}M_{\odot}$) and at the surface.
\smallskip
The time steps were constrained in such a way that the changes in
$log T$, $log \rho$  and $log X_i$ during a time step were smaller
than 0.02, 0.05 and 0.02, respectively. No time steps larger than
$3\times 10^6~y$ were allowed. The implicit difference equations,
replacing the laws of energy and momentum conservation were
iterated
until $log T$ and $log \rho$ were determined within $1\times 10^{-
7}$.
 
\smallskip
\subsection{ Convection }
\smallskip
A region in which $d\rho/dp<0$, or, in the case of uniform
composition and ionization,
a region where the entropy decreased outwards, was considered to
be
convective. In a convective region a complete and instantaneous
mixing was carried out and a convective flux was added to the
radiation flux. The convective flux was calculated according to the
mixing length recipe, using the numerical constant of Mihalas (1973).
 
\smallskip
\subsection{ The Surface Boundary Condition}
\smallskip
The outer half of the last mass shell was treated as a static
atmosphere.
The Stefan-Boltzmann luminosity from the last shell
was equated to the total heat flow into the shell
and the effective temperature was adjusted (by iterations)
until the calculated
entropy (as given by the model atmosphere)
at the base of the atmosphere was equal to the entropy (as given
by the evolution program) at
the center of the shell. The Burlish-Stoer extrapolation method
(Press et al. 1989) was used for the inward integration.
A fine grid of model stellar atmospheres is constructed and the
conditions at the middle of the
last mesh point are found by interpolation inside the
this grid. The model atmosphere start with the photosphere $\tau=2/3$ and
no $T-\tau$ relation is required. The equations of the gray atmosphere
are integrated down to the mass of the prescribed (half) last mass shell.
\smallskip
\subsection{ Diffusion}
\smallskip
Diffusion, caused by density, temperature, pressure,
chemical composition and gravitational potential gradients may play
an
important role in the sun. The effects of helium diffusion on
solar models were first studied by Noerdeling (1977).
Improved calculations were done by Cox et al. (1989), Proffitt and
Michaud (1991), and by Bahcall and Pinsonneault (1992) who
incorporated the approximate analytical description of hydrogen
and helium diffusion by Bahcall \& Loeb (1990) in their standard
solar model calculations. Proffit (1994) and Kovetz \& Shaviv (1994)
have studied the effects of diffusion on the sun by solving
numerically the diffusion equations for all the elements with mass
fraction greater than a prescribed value (usually $10^{-5}$). To simplify
the numerical handling of the ionization, each isotope was assumed to be
in a single ionization stage at each point, the one which is the most
abundant. Consequently, the ionization stage changed from one shell to
the other but in each shell only one state was allowed.
The Kovetz-Shaviv code calculates the diffusion of all the individual
elements from the premain sequence phase to the present age. The
binary
and thermal diffusion coefficients depend on the squared ionic
charges.
The ionization state of each element in every shell is
calculated by solving the Saha equations for all the
elements in each shell. All elements with mass fractions less than
$10^{-5}$ are treated as trace elements. By trace element we mean that
 collisions with other
trace elements are neglected.
\smallskip
As a consequence of diffusion the surface and internal element
abundances change for each element in a slightly different way.
Diffusion
depletes the surface abundances of $^4$He and the heavy elements.
For C, N and O only their present photospheric abundances are
known
(see chapter 3.3.2)  Consequently, in our solar model calculations
their initial solar abundances were adjusted to
reproduce their observed photospheric abundances.
Thus, their initial abundances  should be considered as a prediction
of the standard solar model.
For all other elements the initial meteoritic abundances are used
to predict their final surface abundances which can be compared
with their observed photospheric abundances.
The unknown initial abundance of $^4$He is treated as a free
parameter.
It is adjusted to best reproduce the presently observed sun.
Its predicted surface abundance today can be compared with the
value
derived from helioseismology (Hernandez and Christensen -
Dalsgaard
1994).
 
\bigskip
\section{ Results and Comparison With Observations}
\bigskip
Our predictions of the solar neutrino fluxes for three standard
solar models are summarized in Table IIIa  and are compared
with the results from the four solar neutrino experiments.
The models were calculated with the stellar evolution code of
Kovetz and Shaviv (1994) as described in section 5,
with the improved astrophysical and physical input data which were
described in sections 3 and 4. The three models differ only
in their treatment of element diffusion and equation of state: The model
labeled DSND does not include element diffusion. The models
labeled DS94 and DS95 include element diffusion while the model DS95
includes an improved equation of state for $\Gamma \le 1$. Model
DS94 assumes that the initial heavy metal abundances in the sun
were equal to their meteoritic values, and those of C, N and O
were equal to their observed photospheric abundances, as
summarized
in Table I. In model DS95 the initial abundances of C, N and O were
adjusted to yield their
present day photospheric abundances while the heavy metal
abundances
were assumed to be those found in primitive meteorites, as
summarized in Table I. Table IIIb presents some physical
characteristics of the three models.
The indexes have the following meanings:  c means the core, s the
surface
and 0 means the initial values. ${R}_{conv}$ is the radius of the base
of
the convective envelope and ${T}_{bc}$ is the temperature at the
base of
the convective envelope.
As can be seen from Table IIIa, all three
models yield a $^8$B solar neutrino flux consistent with that
measured
by Kamiokande. However all three models predict capture rates
in the Chlorine and Gallium experiments which are significantly
larger
than those measured by Homestake, GALLEX and SAGE
(We have used the neutrino  cross sections from Table 8.2  of
Bahcall (1989) to convert solar neutrino fluxes to capture rates).
\smallskip
Tables IV present a comparison between four solar models. The
model
labeled BP92 is the best model of BP92. It includes diffusion of
protons and $^4$He but not of other elements. The model
labeled BP95 is the best model of BP95 which includes also diffusion
of the heavy metals but assumes that all the heavy elements diffuse
like fully ionized iron. The predictions of the DS models differ
significantly from the BP models because they differ in
input physics, approximations and numerical methods. Most of the
differences are due to the use of different reaction rates
as summarized in Table II, the different treatments of diffusion and the
equation of state.
This is demonstrated in Table V where we present a comparison
between the best model of BP95 without diffusion labeled BP95ND,
and a solar model, labeled DS(BPND) calculated with the Kovetz-
Shaviv
stellar evolution code with the same physical and astrophysical input
parameters and the same nuclear reaction rates used in BP95ND
and without inclusion
of element diffusion. As can be seen from Tables Va and Vb the two
calculations yield similar results. Even the fluxes  of $^8$B and CNO
solar neutrinos which, under the imposed solar boundary conditions,
are very sensitive to the central solar temperature
differ by less than 4\%.
These remaining  differences are mainly due to the use of
different equations of state, numerical methods,
fine zoning and time steps in the two codes and
due to the inclusion of premain sequence evolution in our code.
\smallskip
To emphasize the important role that might be played by diffusion,
Tables Va and Vb also include  the current best solar models of Dar
and
Shaviv (1995) and of Bahcall and Pinsonneault (1995), which include
diffusion of all elements. As can be seen from Tables Va, Vb, Bahcall
and Pinsonneault (1995) found  rather large increases in their
predicted
$^7$Be, $^8$B, $^{13}$N, $^{15}$O and $^{17}$F
solar neutrino fluxes; 14\%, 36\%, 52\%, 58\%, and 61\%,
respectively, compared with their model (BP95ND) with no diffusion.
These induce 36\%, 33\%, 9\% increases in the predicted
rates in Kamiokande, Homestake, and in GALLEX and SAGE,
respectively.
However, we predict more moderate increases due to diffusion,
4\%, 10\%, 23\%, 24\% and 25\%, respectively, in the above fluxes,
which correspond to 10\%, 10\% and 2\% increases in the predicted
rates in Kamiokande, Homestake, and in GALLEX and SAGE,
respectively.
The differences in the effects of diffusion in DS94 and BP95
are mainly due to two
reasons: (a) In the calculations of Bahcall and Pinsonneault (BP95)
all heavy elements were assumed to diffuse at the same rate as
fully ionized iron while the Dar-Shaviv calculations (DS95)
followed the diffusion of all the elements separately and
used diffusion coefficients for the actual ionization state of each
element.
(b) Bahcall and Pinsonneault assumed that the meteoritic
abundances represent the solar surface abundances today and not
their initial values. They adjusted their initial values to
reproduce  surface abundances today equal to the meteoritic values.
Thus they have actually used an initial ratio
$Z/X=0.0285$ (see Table IV of BP95)
while the observed ratio in meteorites is $Z/X= 0.0245$ (Grevesse
and Noels 1993). Dar and Shaviv (1994; 1995) used the meteoritic values
for the initial metallic abundances and predicted present day
depleted surface abundances. Unfortunately, the uncertainties in the
measured photospheric CNO abundances (typically 30\%-40\%) are
much
larger than their predicted depletion (typically 10\%) and do not
allow a reliable test of this prediction.
 
\smallskip
 
The photospheric abundances of $^7$Li, $^9$Be
and B are much smaller than their abundances in primitive
meteorites.
For instance, the photospheric abundance of $^7$Li is smaller
by nearly a factor of 150 (Anders and Grevesse 1989).
These differences are not explained by the standard solar models
(see e.g., Morrison 1992), in spite of the fact that
significant Lithium burning takes place
during the Hayashi phase. However, observations
of $^7$L in young solar mass stars (e.g., Soderblom et al. 1993)
and solar evolution calculations (e.g. Kovetz and Shaviv 1994)
suggest that premain sequence destruction of $^7$Li together
with its main sequence
destruction is less than a factor 5 and cannot explain a surface
depletion of $^7$Li by nearly a factor 150. Its destruction (via
$^7$Li+p$\rightarrow ^4$He+$^4$He) requires higher temperatures
than  $\sim 2.2\times 10^6K$ predicted by the standard solar models
for the bottom of the convective zone.
Differential rotation below the base of the convective
zone may cause mixing and bring $^7$Li much deeper where it can
burn.
(see, e.g., Vauclair and Charbonnel).
Such mixing may , however, inhibit the inward diffusion of
the heavy elements. Thus, it is not clear whether solar models with
diffusion (such as DS95) provide a more realistic description of the
sun than solar models without diffusion (such as DSND).
Helioseismology data, however, is better explained by SSM's with
diffusion. See, e.g., Cristensen-Dalsgaard et al. 1993.
 
\vfill
\clearpage
\bigskip
\section{Where Are The $^7$Be Neutrinos ? }
\smallskip
 
As we have seen in chapter 6, standard solar models predict capture
rates in the Chlorine and Gallium experiments which are significantly
larger than those measured by Homestake, GALLEX and SAGE.
Combined
with the results from Kamiokande, they seem to suggest that
the $^7$Be solar neutrino flux is strongly suppressed (and perhaps
also
the fluxes of $^8$B and CNO solar neutrinos).
Can standard physics explain the suppression of the $^7$Be solar
neutrino signal in $^{37}$Cl and $^{71}$Ga ? We think that
such a possibility has not been ruled out yet:
Electron capture by $^7$Be into the ground state of $^7$Li
produces 862 keV neutrinos. The threshold
energy for neutrino absorption by $^{37}$Cl is 814 keV. Thus,
absorption
of $^7$Be neutrinos by $^{37}$Cl produces 48 KeV electrons.
The average energy of the pp solar neutrinos is 265 KeV. The
threshold energy for neutrino absorption in $^{71}$Ga is 233 KeV.
Consequently, the produced electron has a typical energy of
33 keV. The de Broglie wave lengths of such electrons are
larger than the Bohr radii of the atomic K shells in Cl and Ga
and their energies are similar to the kinetic energies of electrons
in the K shells. Consequently, screening of the nuclear
charge by atomic electrons and final state interactions (exchange
effects, radiative corrections, nuclear recoil against the electronic
cloud, etc.) may reduce the absorption cross sections of
pp neutrinos in $^{71}$Ga (perhaps making room for the expected
contribution of $^7$Be in Gallium ?)
and of $^7$Be neutrinos in $^{37}$Cl (perhaps making the solar
neutrino
observations of Kamiokande and the Homestake experiment
compatible
Dar, Shaviv and Shaviv 1996). It is interesting
to note that although final state interactions in Tritium beta
decay have been studied extensively, they do not explain yet why its
end-point spectrum ($E_e\sim 18.6~kEV$)
yields consistently, in all recent measurements,
a negative value for the squared mass of the electron neutrino.
Final state interactions in $^{37}$Cl and $^{71}$Ga are expected
to be much larger because of their much larger values of Z.
Note also that this explanation of the solar neutrino problem
implies that experiments such as
BOREXINO and HELLAZ will observe the full $^7$Be solar neutrino
flux
while the MSW solution predicts that it will be strongly suppressed.
 
Even if the $^7$Be solar neutrino flux is strongly suppressed,
it does not eliminate yet standard physics solutions to the
solar neutrino problem. For instance,
collective plasma physics effects, such as
very strong magnetic or electric fields near the center of the sun,
may polarize  the plasma electrons, and affect the
branching ratios of electron capture by $^7$Be (spin $3/2^-$) into
the ground state
(spin $3/2^-$,  $E_{\nu_e}=0.863~MeV$, BR=90\% and the excited
state
(spin $1/2^-$,  $E_{\nu_e}=0.381~MeV$, BR=10\%) of $^7$Li. Since
solar neutrinos with $E_{\nu_e}=0.381~MeV$ are below the
threshold (0.81
MeV) for capture in $^{37}$Cl and have a capture cross section in
$^{71}$Ga that is smaller by about a factor of 6 relative to
solar neutrinos with $E_{\nu_e}=0.863~MeV$, therefore a large
suppression in the branching ratio to the ground state can produce
large suppressions of the $^7$Be solar neutrino signals in $^{37}$Cl
and in $^{71}$Ga.

\bigskip
\section {The MSW Solution}
\bigskip
Standard solar models, like the one presented in this work,
perhaps can explain the results reported by
Kamiokande. However, standard physics cannot explain an $^{37}$Ar
production rate in $^{37}$Cl smaller than that expected from
the solar $^8$B neutrino flux measured by Kamiokande. If the
experimental results of Kamiokande and Homestake are interpreted
as
an evidence for such a situation (e.g., Bahcall 1994; 1995),
they do imply  new physics beyond the standard particle physics
model
(Bahcall and Bethe 1991). In that case
an elegant solution to the solar neutrino anomaly is
resonant neutrino flavor conversion in the sun, first proposed
by Mikheyev and Smirnov (1986) (see also Wolfenstein 1978; 1979).
Many authors have carried out extensive calculations to determine
the neutrino mixing parameters
which can bridge between the predictions of the
standard solar models and the solar neutrino observations. They
found
that  a neutrino mass difference
$\Delta m^2\sim 0.7\times 10^{-5}~eV^2 $ and a neutrino mixing of
$sin^2 2\theta\approx 0.5\times 10^{-2}$ can solve the solar
neutrino
problem (see, e.g., Krastev and Petcov 1993; Castellani et al. 1993;
Hata et al. 1994; Kwong and Rosen 1994; Berezinsky et al. 1994). In
fact,
these parameters can be found from simple analytical considerations:
\smallskip
As we have seen, the results  of  GALLEX and SAGE can be predicted
directly from the pp flux, which is fixed by the solar luminosity,
and the $^8$B flux measured by Kamiokande, if GALLEX and SAGE
are blind to all other neutrinos. This blindness may be due
to flavor conversion of these neutrinos.
The resonance condition for flavor conversion in the sun is
\begin{equation}
n_e=n^{res}_e={\Delta m^2 cos 2\theta\over 2\sqrt 2G_FE_\nu}
\approx {\Delta m^2\over 2\sqrt{2}G_FE_\nu},
\end{equation}
for small mixing angles. Solar pp neutrinos, whose maximum energy
is 0.42 MeV do not encounter  resonance density if the resonance
density for $E_\nu=0.42$ is larger than the central electron density
in the sun. The 0.861 MeV $^7$Be solar neutrinos, and all the other
solar neutrinos which are more energetic, encounter a resonance
density
in the sun and suffer a resonance flavor conversion if the resonance
condition is already satisfied for 0.861 MeV neutrinos. The electron
density at the center of the sun is $n_e\approx 6\times 10^{25}
~cm^{-3}$ and
the last two conditions together with Eq. 21 yield
\begin{equation}
 0.5\times 10^6~eV^2\leq \Delta m^2 \leq 1\times 10^6~eV^2.
\end{equation}
The probability of flavor conservation is given approximately by
(e.g. Parke 1986, Dar et al. 1987)
$P(\nu_e\rightarrow\nu_e)\approx e^{-\epsilon/E_\nu},$ where
\begin{equation}
\epsilon={\pi H\Delta m^2 sin^2\theta\over 4 cos 2\theta}\approx
           {\pi H\Delta m^2 \theta^2\over 4},
\end{equation}
where $H$ is the solar scale height at the resonance.
Thus, we see that the suppression of the  solar $\nu_e$ flux
can still be adjusted by the choice of the mixing angle $\theta$.
A suppression factor of $P\approx e^{-1}$ at the most effective solar
neutrino energies ($\sim 10~MeV$) in the chlorine detector
yields a suppression factor of $e^{-10}$ at $E_\nu\sim 1~MeV. $
Such a suppression and $\Delta m^2 \approx 0.75 \times 10^{-
5}~eV^2$
yield a mixing parameter $sin^2 2\theta\approx  0.5\times 10^{-2}$.
\smallskip
 
Are neutrino oscillations responsible for the solar neutrino anomaly?
The answer  may be provided by two experiments planned to begin
collecting data in 1966:
\smallskip
The Sudbury Neutrino Observatory (SNO) will detect two neutrino
interactions in 1000 tons of heavy water:
 
\begin{equation}
\nu_e +D\rightarrow p+p+e^{-}~,
\end{equation}
and
\begin{equation}
\nu_x+D\rightarrow p+n+\nu_{x}~.
\end{equation}
 
If more solar neutrinos are detected via the second reaction (which
is blind to the neutrino flavor) than
via the first reaction (where only $\nu_e$'s can contribute),
it would provide evidence that some solar $\nu_e$'s have changed
their flavor before reaching the detector. The total solar
neutrino flux will also be measured at SNO by elastic scattering
on electrons in the light water shield. The expected counting rate in
the 1000 tones of heavy water of SNO is about 10 per day!.
\smallskip
The Superkamiokande experiment will utilize 50,000 tones of
high purity light water and will enhance the counting rate of
solar neutrinos in Kamiokande by about a factor 30.
It should have enough statistics to detect the
change in the energy spectrum of the high energy $^8$B solar
neutrinos predicted by the MSW (Mikheyev and Smirnov 1986,
Wolfenstein
1978) solution to the solar neutrino problem.
\bigskip
\section{ Summary and Conclusions}
\bigskip
The results of the four pioneering solar neutrino experiments,
confirm that the sun derives its energy by
fusion of hydrogen into helium in its core. This is
a great triumph both for experimental and theoretical physics.
The reliability of the results from these difficult and ingenious
experiments is supported by their consistency,
and recently, most convincingly, by the
GALLEX Chromium source experiment .
\smallskip
The capture rates of solar neutrinos measured by GALLEX and SAGE
are
above the minimal rates expected from the solar luminosity
and the conservation of lepton flavor. The
$^8$B solar neutrino flux predicted by our improved standard solar
model, which was described in this paper, is consistent, within the
theoretical and experimental uncertainties, with the solar neutrino
observations at Homestake and Kamiokande. However,
the experimental results from the four solar neutrino
experiments seem to suggest that the flux of $^7$Be solar
neutrinos at Earth is much smaller than that predicted by the
standard solar models, including our improved SSM
which includes premain sequence evolution, element diffusion,
partial ionization effects, all possible nuclear reactions
between the main elements, uses updated values for the initial
solar element abundances, the solar age, the solar luminosity,
the nuclear reaction rates and the radiative opacities and does not
impose either nuclear equilibrium or complete ionization.
\smallskip
Solutions to the solar neutrino problem which
do not invoke neutrino properties beyond the standard electroweak
model
are not ruled out yet:
 
The solar neutrino problem may be a terrestrial
problem. The neutrino capture cross sections near threshold
in the radiochemical experiments may be different
from the calculated cross sections (Dar, Shaviv and Shaviv 1996).
The inferred solar neutrino fluxes
from the GALLEX and HOMESTAKE experiments may be different
from
the true solar neutrino fluxes. They do
not establish beyond doubt that there is a real $^7$Be solar
neutrino
deficit. Perhaps, future experiments such as
BOREXINO and HELLAZ  will be able to establish
that.

The solar neutrino problem may be an astrophysical problem.
The deviations of the experimental results from those predicted by
the
standard solar models may reflect the approximate
nature of the standard solar models (which neglect, or treat
only approximately, many effects and do
not explain yet solar activity nor the surface depletion of
Lithium, Berilium and Boron relative to their meteoritic values,
which may or may not be relevant to the solar neutrino problem).
Improvements of the standard solar model should continue.
In particular,
dense plasma effects on nuclear reaction rates and radiative
opacities,
which are not well understood, may strongly affect the SSM
predictions
and should be further studied, both theoretically and experimentally.
Relevant information may be obtained from studies of thermonuclear
plasmas in inertial confinement experiments. Useful information
may also be obtained from improved data on screening effects
in low energy nuclear cross sections of ions, atomic beams and
molecular
beams incident on a variety of gas, solid and plasma targets.
 
Better knowledge of low energy nuclear cross sections is
badly needed. Measurement of crucial low energy nuclear cross
sections
by new methods, such as measurements of the
cross sections for the radiative captures ${\rm p+^7Be\rightarrow
^8B+\gamma}$ and ${\rm ^3He+^4He\rightarrow ^7Be+\gamma}$ by
photodissociation of $^8$B  and $^7$Be in the coulomb field of
heavy nuclei are highly desirable.

The $^{37}$Ar production rate in $^{37}$Cl may indeed be smaller
than that expected from the flux of standard solar neutrinos as
measured
by electron scattering in the Kamiokande experiment. In that case
neutrino oscillations, and in particular the MSW effect, may
be the correct solution to the solar neutrino problem. Only future
experiments, such as SNO, Superkamiokande, BOREXINO and HELLAZ,
will be able to supply
a definite proof that Nature has made use of this beautiful effect.
\smallskip
\section{Acknowledgment}
The authors would like to thank an anonymous referee for very useful comments.
This research was supported in part by the
Technion Fund For Promotion of Research.
\vfill
\eject
\bigskip
\section{ REFERENCES}
\smallskip

 \noindent
Abdurashitov, J.N. et al., 1995, Nucl. Phys. B (Proc. Suppl.)
{\bf 38}, 60.

 \noindent
Anders, E. and Grevesse, N., 1989, Geochim. Cosmochim. Acta,
{\bf 53}, 197.

 \noindent
Anselmann, P. et al., 1995a,
Nucl. Phys. B (Proc. Suppl.) {\bf 38}, 68.

 \noindent
Anselmann, P. et al., 1995b,  Phys. Lett. B {\bf 342}, 440.

 \noindent
Bahcall, J.N. 1989, "Neutrino Astrophysics", (Cambridge
University Press 1989).

 \noindent
Bahcall, J.N., 1994, Phys. Lett. B {\bf 338}, 276.

 \noindent
Bahcall, J.N., 1995, Nucl. Phys. B (Proc. Suppl.) {\bf 38}, 98.

 \noindent
Bahcall, J.N. \& Bethe, H. 1991, Phys. Rev. D. {\bf 44}, 2962.

 \noindent
Bahcall, J.N. \& May, R.M., 1968, ApJ. {\bf 155}, 501.

 \noindent
Bahcall, J.N. \& Pinsonneault, M. 1992, Rev. Mod. Phys. {\bf 64}, 885.

 \noindent
Bahcall, J.N. \& Pinsonneault, M. 1995, Rev. Mod. Phys., (submitted).

 \noindent
Bahcall, J.N. and Ulrich, R.K., 1988, Rev. Mod. Phys. {\bf 60}, 297.
 
  \noindent
Barish, B.C., 1995, Nucl. Phys. B (Proc. Suppl) {\bf 38} 343
 
  \noindent
Barker, F.C. \& Spear, R.H. 1986, ApJ. {\bf 307}, 1986
 
  \noindent
Berezinsky, V., 1994, Comm. Nucl. Part. Phys. {\bf 21}, 249
 
  \noindent
Berezinsky, V., et al, 1994, Phys. Lett. {\bf 341}, 38
 
  \noindent
Bethe, H. 1939, Phys. Rev. {\bf 55}, 103
 
  \noindent
Castellani, V. et al., 1994, Phys. Lett. B {\bf 324}, 425
 
  \noindent
Caughlan, G.R. \& Fowler, W.A. 1988, Atomic and Nucl. Data
Tables {\bf 40}, 283
 
  \noindent
Christensen-Dalsgaard, J., 1994, Europhys. News {\bf 25}, 71
 
  \noindent
Christensen-Dalsgaard, J. \&
Dappen, W. 1992, A\&A Rev. {\bf 4}, 267
 
  \noindent
Christensen-Dalsgaard, J. et al., 1993, ApJ. {\bf 403}, L75
 
  \noindent
Clayton, D. 1968, Principles of Stellar Evolution \& Nucleosyn.
(McGraw-Hill)
 
  \noindent
Cleveland, B.T. et al., 1995, Nucl. Phys. B (Proc. Suppl.) {\bf 38}, 47
 
  \noindent
Cox, A.N., Guzik, J.A. \& Kidman, R.B. 1989, ApJ. {\bf 342}, 1187
 
  \noindent
Dar, A., 1993, in {\it Particles and Cosmology},
(World Scientific, Eds. E.N. Alexyev et al.) p.3
 
  \noindent
Dar, A. et al., 1987, Phys. Rev. D {\bf 35}, 3607
 
  \noindent
Dar, A. and Nussinov, S. 1991, Particle World {\bf 2}, 117
 
  \noindent
Dar, A. \& Shaviv G. 1994, Proc. VI Int. Conf. on Neutrino Telescopes
(edt. M. Baldo-Ceolin) p. 303. See also Shaviv G., 1994
 
  \noindent
Dar, A., Shaviv, G. \& Shaviv, N., 1996, to be published
 
  \noindent
Dawarakanath, M.R., \& Winkler H.,  1971, Phys. Rev. C {\bf 4}, 1532
 
  \noindent
Descouvemont and Baye, 1994, Nucl. Phys. {\bf A567}, 341
 
  \noindent
Dzitko, H. et al., 1995, ApJ. {\bf 447}, 428 (1995)
 
  \noindent
Engstler, S. et al, 1988, Phys. Lett. B {\bf 202}, 179
 
  \noindent
Filippone, B.W. 1986, Ann. Rev. Nucl. Sci. {\bf 36}, 717
 
  \noindent
Filippone, B.W. et al. 1983, Phys. Rev. C {\bf 28}, 2222
 
  \noindent
Fogli, G.J. and Lisi. E, (1995) Astroparticle Phys. {\bf 3}, 185
 
  \noindent
Gai, M., 1995,  Nucl. Phys. B (Proc. Suppl.) {\bf 38}, 77
 
  \noindent
Geiss, J.,  1993  in Origin and Evolution of the Elements, ed.
N. Prantzos et al (Cambridge Univ. Press, Cambridge) p. 89
 
  \noindent
Goldsmith, S. et al., 1984, Phys. Rev. {\bf A30}, 2775
 
  \noindent
Gopel, C. et al., 1994, Earth Sci. Lett.
 
  \noindent
Greife, U. et al., 1994, Nucl. Inst. \& Methods. A {\bf 350}, 327
 
  \noindent
Grevesse, N., 1991, A\&A, {\bf 242}, 488
 
  \noindent
Grevesse, N. \& Noels, A., 1993, in {\it Origin and Evolution of the
Elements, eds.. Prantzos et al.} (Cambridge Univ. Press) p. 15
 
  \noindent
Grevesse, N. \& Noels, A., 1993, Phys. Scripta {\bf T47} , 133
 
  \noindent
Guenter, D.B. 1989, ApJ. {\bf 339}, 1156
 
  \noindent
Hata, N. et al., 1994, Phys. Rev. D {\bf 49}, 3622
 
  \noindent
Hata, N. \& Langacker, P., (1995), Phys. Rev. D {\bf 52}, 420
 
  \noindent
Hernandez, E.P. \& Christensen-Dalsgaard, J., (1994), MNRAS
{\bf 269}, 475
 
  \noindent
Hickey, J.R., 1982, J. Solar Energy {\bf 29}, 125
 
  \noindent
Hilgemeier, M. et al., 1988, Z. Phys. A {\bf 329}, 243
 
  \noindent
Johnson, C.W. et al. 1992, ApJ. {\bf 392}, 320
 
  \noindent
Kajino, T., Toki, H. \& Austin, S.M. 1987, ApJ. {\bf 319}, 531
 
  \noindent
Kajita, 1994, ICRR-Report 332-94-27 (December 1994).
 
  \noindent
Kamionkowski, M. \& Bahcall, J., (1994), ApJ. {\bf 420}, 884
 
  \noindent
Kavanagh, R.W., et al., 1969, Bull. Am. Phys. Soc. {\bf 14}, 1209
 
  \noindent
Kim, Y.E., et al., 1995,  Nucl. Phys. B (Proc. Suppl.) {\bf 38}, 293
 
  \noindent
Kovetz, A. \& Shaviv, G., 1994, ApJ. {\bf 426}, 787
 
  \noindent
Krastev, P.I. \& Petcov, S.T. 1993, Phys. Lett. B {\bf 299}, 99
 
\noindent
Krauss, A. et al., 1987, Nucl. Phys. A {\bf 467}, 273
 
  \noindent
Krawinkel, H. et al., 1982, Z. Phys. A, {\bf 304}, 307
 
  \noindent
Kubono,  S. et al., 1989, {\bf 344}, 460
 
  \noindent
Kwong, W. \& Rosen, S, 1994, Phys. Rev. Lett. {\bf 73}, 369
 
  \noindent
Langacker, P., 1995, Nucl. Phys. B (Proc. Suppl.) {\bf 38}, 152
 
  \noindent
Langanke, K. 1991, in Nuclei in The Cosmos (ed. H, Oberhummer)
Berlin:
Springer-Verlag, p. 61
 
  \noindent
Lee, R.B. et al., 1991, Metrologica {\bf 28}, 265
 
  \noindent
Linsky, J.L. et al., 1993, ApJ. {\bf 402}, 694
 
  \noindent
Louis, W.C., 1995, Nucl. Phys. B (Proc. Suppl.) {\bf 38}, 229
 
  \noindent
Mihalas, D. 1978, Stellar Atmospheres, W.H. Freeman \& Co., p. 88
 
  \noindent
Mikheyev, P. \& Smirnov, A. Yu. 1985, Yad. Fiz. {\bf 42}, 1441
 
  \noindent
Motobayashi, T., et al., 1994, Phys. Rev. Lett., {\bf 73}, 2680
 
  \noindent
Nagatani, K. et al., 1969, Nucl. Phys. {\bf A 128}, 325
 
  \noindent
Osborne, J.L. et al., 1984, Nucl. Phys. {\bf A419}, 115
 
  \noindent
Parke, S.J. 1995, Phys. Rev. Lett. {\bf 74}, 839
 
  \noindent
Parke, S.J. 1986, Phys. Rev. Lett. {\bf 57}, 1275
 
  \noindent
Parker, P.D., 1966, Phys. Rev. Lett. {\bf 150}, 851
 
  \noindent
Parker, P.D., 1968, ApJ. {\bf 153}, L85
 
  \noindent
Parker, P.D., \& Kavanagh, R. W., 1963 Phys. Rev. {\bf 131}, 2578
 
  \noindent
Press, W.H. et al., 1989, Numerical Recipes, Cambridge Univ. Press,
p. 563
 
  \noindent
Proffitt, C.R. \& Michaud, G., 1991, ApJ. {\bf 380}, 238
 
  \noindent
Proffitt, C.R., 1994, ApJ. {\bf 425}, 849
 
  \noindent
Robertson, R.G.H. 1972, Phys. Rev. C {\bf 7}, 543
 
  \noindent
Robertson, R.G.H. et al., 1983, Phys. Rev. C {\bf 27}, 11
 
  \noindent
Rogers, F.J. \& Iglesias, C.A., 1992, ApJ. Suppl. {\bf 79}, 57
 
  \noindent
Rolfs, C., 1994  Nucl. Phys. B (Proc. Suppl.) {\bf 35} 334
 
  \noindent
Sackman, I.J. et al., 1990, ApJ. {\bf 360}, 727
 
  \noindent
Schramm, D.N. \& Shi, X, 1994, Nucl. Phys. B (Proc. Suppl.){\bf 35},
321
 
  \noindent
Shaviv, G. 1995, Nucl. Phys. B (Proc. Suppl.) {\bf 38}, 81
 
  \noindent
Shaviv, G. \& Shaviv, N., 1995, to be published
 
  \noindent
Shi, X.D. et al., 1994, Phys. Rev. D {\bf 50}, 2414
 
  \noindent
Shoppa, T.D. et al., 1993, Phys. Rev. C {\bf 48}, 837
 
  \noindent
Soderblom, D.R. et al., 1994
 
  \noindent
Sturenburg, S. \& Holweger, H., 1990, A\&A {\bf 237}, 125
 
  \noindent
Suzuki, Y. 1995, Nucl. Phys. B (Proc. Suppl.) {\bf 38}, 54
 
  \noindent
Typel, S. et al. 1991, Z. Phys. A {\bf 339}, 249
 
  \noindent
Tilton, G.R., 1988, in {\it Meteorites \& the Early Solar System},
(eds. J.F. Kerridge \& M.S. Matthews, Univ. of Arizona Press 1988)
p. 259
 
  \noindent
Vauclair, S \& Charbonnel, C., 1995, A\&A, {\bf 295}, 715
 
  \noindent
Turck-Chieze, S. et al., 1988, ApJ. {\bf 335}, 415
 
  \noindent
Turck-Chieze, S. \& Lopes, I., 1993, ApJ. {\bf 408}, 347
 
  \noindent
Vaughn et al., 1970, Phys. Rev. {\bf C2}, 1657
 
  \noindent
Volk, H. et al., 1983, Z. Phys. A {\bf 310}, 91
 
  \noindent
Wasserburg, G.J., et al., 1995, Ap. J. in Press
 
  \noindent
Wiescher, M.  et al., 1989, ApJ, {\bf 343}, 352
 
  \noindent
Wilson, R.C. \& Hudson, H.S., 1988, Nature {\bf 332}, 810
 
  \noindent
Wilson, R.C., 1993, Atlas of Satellite Observations (eds. R.J. Gurney
et al. Cambridge Univ. Press, NY) p. 5
 
  \noindent
Wolfenstein, L. 1978, Phys. Rev. {\bf D17}, 2369
 
  \noindent
Wolfenstein, L. 1979, Phys. Rev. {\bf D20}, 2634
 
  \noindent
Xu et al., (1994),  Phys. Rev. Lett. {\bf 73}, 2027
 
\clearpage
{\bf Table I:}  Summary of Information
on Abundances of Various Elements relative to Hydrogen
( A$\equiv$log([A]/[H])+12) in  Primitive Meteorites, in the Solar
Photosphere, in the Solar Wind and in the Local Interstellar Medium,
Used In The DS Standard Solar Models.
\vfill
\eject
{$$\matrix{{\rm Element}&{\rm Abundance}&{\rm Source}&{\rm
Reference}\cr
{\rm D }&7.22\pm 0.05&{\rm Meteorites,~Solar Wind} & {\rm
Linsky~1993,~
Geiss~1993}\cr
^{\rm 3}{\rm He}&7.18\pm0.08&{\rm Meteorites,~Solar~Wind}&{\rm
Geiss~1993}\cr
^{\rm 7}{\rm Li}&1.54\pm
0.0X&{\rm Meteorites} & {\rm Anders~and~Grevesse~1989}\cr
^{\rm 9}{\rm Be}&1.13\pm
0.0X&{\rm Meteorites} & {\rm  Anders~and~Grevesse~1989}\cr
^{\rm 12}{\rm C}&8.55\pm
0.05&{\rm Photosphere} & {\rm Grevesse~and~Noels~1993}\cr
^{\rm 13}{\rm C}&6.60\pm
0.05& {\rm Photosphere} & {\rm Grevesse~and~Noels~1993}\cr
^{\rm 14}{\rm N}&7.97\pm 0.07&
{\rm Photosphere }& {\rm Grevesse~and~Noels~1993}\cr
^{\rm 16}{\rm O}&8.78\pm 0.07&
{\rm Photosphere } & {\rm Grevesse~and~Noels~1993}\cr
^{\rm 20}{\rm Ne}&8.08\pm 0.06&
{\rm Photosphere } & {\rm Grevesse~and~Noels~1993}\cr
^{\rm 23}{\rm Na}&6.33\pm 0.03&
{\rm Meteorites~and~Photosphere } & {\rm
Grevesse~and~Noels~1993}\cr
^{\rm 24}{\rm Mg}&7.58\pm 0.05&" &"\cr
^{\rm 27}{\rm Al }&6.47\pm 0.07&"&"\cr
^{\rm 28}{\rm Si}&7.66\pm 0.05&"&"\cr
^{\rm 31}{\rm P}&5.45\pm 0.04&"&"\cr
^{\rm 32}{\rm S}&7.21\pm 0.06&"&"\cr
^{\rm 35}{\rm Cl }&5.5\pm 0.3&"&"\cr
^{\rm 40}{\rm Ar}&6.52\pm 0.1&"&"\cr
^{\rm 40}{\rm Ca}&6.36\pm 0.02&"&"\cr
^{\rm 40}{\rm K}&4.85&"&"\cr
^{\rm 45}{\rm Sc}&3.08&"&"\cr
^{\rm 48}{\rm Ti}&5.02\pm 0.06&"&"\cr
^{\rm 50}{\rm V}&3.99&"&"\cr
^{\rm 52}{\rm Cr}&5.67\pm 0.03&"&"\cr
^{\rm 55}{\rm Mn}&5.39\pm 0.03&"&"\cr
^{\rm 56}{\rm Fe}&7.50\pm 0.04&"&"\cr
^{\rm 63}{\rm Cu}&4.15&"&"\cr
^{\rm 58}{\rm Ni}&6.25\pm 0.04&"&"\cr
^{\rm 64}{\rm Zn}&4.33&"&"\cr
\rm Z\rm /\rm X&\rm 0.0245&"&"\cr}$$}
 
\bigskip
{\bf Table II:}
Comparison Between the Astrophysical S Factors
for the pp-chain Reactions used in BP95 and in DS94 and DS95.
The values of S are given in $keV\cdot b$ Units.
 \bigskip
 
$$\matrix{ {\rm Reaction } \hfill & {\rm {S}^{BP}}\rm (0) \hfill
&{\rm
S}^{DS}\rm (0)
 \hfill \cr
^{\rm 1}{\rm H}(\rm p\rm ,{\rm e}^{\rm +}{\rm \nu }_{e}{\rm ) D}
\hfill &
\rm 3.896\times {10}^{-22} \hfill &\rm 4.07\times {10}^{-22} \hfill
\cr
^{\rm 1}{\rm H}({\rm p}{e}^{\rm -}{\rm \nu}_{e} ){\rm D} \hfill &
{\rm Bahcall~89 } \hfill &{\rm CF88} \hfill \cr
^{\rm 3}{\rm He}(^{\rm 3}{\rm He},2\rm p\rm ){\rm He}^{\rm 4}
\hfill &
4.99\times {10}^{3} \hfill &5.6\times {10}^{3} \hfill \cr
^{\rm 3}{\rm He}(^{\rm 4}{\rm He},\rm \gamma \rm )^{\rm 7}{\rm
Be}
\hfill &0.524 \hfill & 0.45 \hfill \cr
^{\rm 7}{\rm Be}({\rm e}^{\rm -},{\rm \nu }_{e}{\rm )}{\rm Li}^{\rm
7}
\hfill & {\rm Bahcall~89} \hfill &{\rm CF88} \hfill \cr
^{\rm 7}{\rm Be}(\rm p\rm ,\rm \gamma \rm )^{\rm 8}{\rm B} \hfill
&0.0224 \hfill
& 0.017 \hfill \cr}$$
 
\clearpage
 
{\bf Table IIIa:} Comparison Between Solar Neutrino Fluxes
Predicted by the current best Standard Solar Models of Dar and
Shaviv, with and without
element diffusion, and the Solar Neutrino Observations.

$$\matrix{\nu~{\rm Flux}\hfill & {\rm  DSND} \hfill & {\rm  DS94}
\hfill
& {\rm   DS95}
\hfill &{\rm  Obs} \hfill &{\rm
Exp} \hfill \cr
{\rm \phi }_{\nu }(pp)~[{10}^{10}{cm}^{-2}{s}^{-1}] \hfill &6.10
\hfill &6.06 \hfill &6.10 \hfill  &\mit &\rm \cr
{\phi }_{\nu }(pep)~[{10}^{8}{cm}^{-2}{s}^{-1}] \hfill &
1.43\hfill &1.42\hfill&1.43\hfill &\rm \hfill &\rm \cr
{\phi }_{\nu }(^{7}{Be})~[{10}^{9}{cm}^{-2}{s}^{-1}]\hfill &4.03\hfill
&4.00\hfill &3.71\hfill &\rm \ll \phi_{\nu}^{SSM}(^7Be)\hfill
&\rm ALL\cr
{\phi }_{\nu }(^{8}{B})~[{10}^{6}{cm}^{-2}{s}^{-1}]\hfill &2.54\hfill
&2.60\hfill & 2.49 \hfill &\rm 2.9\pm 0.4\hfill &\rm Kam.\cr
{\phi }_{\nu }(^{13}{N})~[{10}^{8}{cm}^{-2}{s}^{-1}]\hfill &3.21\hfill
&3.30\hfill
& 3.82 \hfill &\rm \hfill &\rm \cr
{\phi }_{\nu }(^{15}{O})~[{10}^{8}{cm}^{-2}{s}^{-1}]\hfill &3.13\hfill
&3.19\hfill
& 3.74 \hfill &\rm \hfill &\rm \cr
{\phi }_{\nu }(^{17}{F})~[{10}^{6}{cm}^{-2}{s}^{-1}]\hfill &3.77\hfill
&3.84\hfill
& 4.53 \hfill &\rm \hfill &\rm \cr
\Sigma (\phi \sigma)_{Cl}~[SNU]\hfill &\rm 4.2\pm 1.2\hfill &
\rm 4.2\pm 1.2\hfill &\rm 4.1\pm 1.2\hfill &\rm 2.55\pm 0.25
 \hfill &{\rm Home.} \cr
\Sigma(\phi\sigma)_{Ga}~[SNU]\hfill &\rm 116\pm 6\hfill
&\rm 116\pm 6 \hfill &\rm 115\pm 6\hfill
&\rm 79\pm 12 \hfill &{\rm GALLEX}\cr
\Sigma(\phi\sigma)_{Ga}~[SNU]\hfill &\rm 116\pm 6\hfill
&\rm 116\pm 6 \hfill &\rm 115\pm 6\hfill
&\rm 74\pm 16 \hfill &{\rm SAGE}\cr}$$
 
\vfill
\eject
 
{\bf Table IIIb:} Characteristics of the DS
Solar Models in Table IIIa (c=center; s=surface;
bc=base of convective zone; ${\bar N} =log([N]/[H] + 12)$.
 
$$\matrix{{\rm Parameter}\hfill& {\rm DSND}\hfill& {\rm
DS94}\hfill& {\rm
DS95}\hfill \cr
{T}_{c}~[{10}^{7}K] \hfill &1.553 \hfill &1.554 \hfill
& 1.561 \hfill \cr
{\rho }_{c}~[g~c{m}^{-3}]\hfill&154.9 \hfill&155.3\hfill
& 155.4 \hfill \cr
{X}_{c}\hfill&0.3491\hfill &0.3462\hfill &0.3424 \hfill \cr
{Y}_{c}\hfill&0.6333 \hfill &0.6359 \hfill &0.6380 \hfill \cr
{Z}_{c}\hfill&0.01757 \hfill &0.01802 \hfill
&0.01950 \hfill \cr
{X}_{s}\hfill&0.6978 \hfill &0.7243 \hfill &0.7512 \hfill \cr
{Y}_{s}\hfill&0.2850 \hfill &0.2597 \hfill &0.2308 \hfill \cr
{Z}_{s}\hfill&0.01703 \hfill &0.01574 \hfill
&0.0170 \hfill \cr
\overline{N}_s{(^{12}C})\hfill&8.55\hfill &8.50 \hfill&8.55 \hfill \cr
\overline{N}_s{(^{14}N})\hfill&7.97\hfill &7.92 \hfill&7.97 \hfill \cr
\overline{N}_s{(^{16}O})\hfill&8.87\hfill &8.82 \hfill&8.87 \hfill \cr
\overline{N}_s{(^{20}Ne})\hfill&8.08\hfill &8.03 \hfill&8.08 \hfill \cr
\overline{X(\geq^{24}Mg)/{Z}_s}\hfill&0.00464
\hfill &0.00414 \hfill &0.00415 \hfill \cr
{R}_{conv}~[R/R_{\odot}]\hfill&0.7306 \hfill &0.7105 \hfill &0.7301
\hfill \cr
{T}_{bc}~[{10}^{6}{\rm K}]\hfill&1.97 \hfill &2.10 \hfill &2.105
\hfill \cr
{T}_{eff}~[{\rm K}]\hfill&5895 \hfill &5920 \hfill &5803
\hfill \cr
{X}_{0}\hfill&0.69775\hfill &0.69859 \hfill &0.7295
\hfill \cr
{Y}_{0}\hfill&0.285007 \hfill &0.28267 \hfill &0.2509
\hfill \cr
{Z}_{0}\hfill&0.01703 \hfill & 0.01931\hfill &
0.01833\hfill \cr}$$
 
\vfill
\eject
 
{\bf Table IVa:} Comparison Between Solar Neutrino Fluxes
Predicted by the Dar-Shaviv  Models and the
Bahcall-Pinsonneault best Solar Models.
 
$$\matrix{\rm & {\rm BP92}& {\rm  DS94} & {\rm BP95}& {\rm
DS95}\cr
{\phi }_{\nu }(pp)~[{10}^{10}{cm}^{-2}{s}^{-1}]\hfill &
6.00\hfill &6.06\hfill &5.91\hfill &6.10\cr
{\phi }_{\nu }(pep)~[{10}^{8}{cm}^{-2}{s}^{-1}]\hfill &
1.43\hfill & 1.42\hfill &1.40\hfill &1.43\cr
{\phi }_{\nu }(^{7}{Be})~[{10}^{9}{cm}^{-2}{s}^{-1}]\hfill &
4.89\hfill &4.00\hfill &5.15\hfill &3.71\cr
{\phi }_{\nu }(^{8}{B})~[{10}^{6}{cm}^{-2}{s}^{-1}]\hfill &
5.69\hfill &2.60\hfill &6.62\hfill &2.49\cr
{\phi }_{\nu }(^{13}{N})~[{10}^{8}{cm}^{-2}{s}^{-1}]\hfill &
4.92\hfill &3.30\hfill &6.18\hfill &3.82\cr
{\phi }_{\nu }(^{15}{O})~[{10}^{8}{cm}^{-2}{s}^{-1}]\hfill &
4.26\hfill &3.19\hfill &5.45\hfill &3.74\cr
{\phi }_{\nu }(^{17}{F})~[{10}^{6}{cm}^{-2}{s}^{-1}]\hfill &
5.39\hfill &3.84\hfill &6.48\hfill &4.53\cr
\Sigma (\phi \sigma)_{Cl}~[SNU]\hfill &\rm 8\pm 1\hfill &
\rm 4.2\pm 1.2\hfill
&\rm 9.3\pm 1.4\hfill&\rm 4.1\pm 1.2\hfill \cr
\Sigma(\phi\sigma)_{Ga}~[SNU]\hfill &\rm 132\pm 7\hfill
&\rm 116\pm 6\hfill &\rm 137\pm 8\hfill
&\rm 115\pm 6\hfill \cr}$$
 
\vfill
\eject
{\bf Table IVb} Characteristics of the BP95, DS94, and DS95
Solar Models in Table IIIa
(c=center; s=surface; bc=base of convective zone;
${\rm \bar N=log([N]/[H])+12)}$.
 
$$\matrix{{\rm Parameter}\hfill& {\rm BP95}\hfill& {\rm
DS94}\hfill& {\rm
DS95}\hfill \cr
{T}_{c}~[{10}^{7}K] \hfill &1.584 \hfill &1.554 \hfill &1.561 \hfill \cr
{\rho }_{c}~[g~c{m}^{-3}]\hfill&156.2 \hfill&155.3\hfill&155.4\hfill
\cr
{X}_{c}\hfill&0.3333\hfill &0.3462\hfill &0.3424 \hfill \cr
{Y}_{c}\hfill&0.6456 \hfill &0.6359 \hfill &0.6380 \hfill \cr
{Z}_{c}\hfill&0.0211\hfill&0.01950 \hfill
&0.01940 \hfill \cr
{X}_{s}\hfill&0.7351 \hfill &0.7243 \hfill &0.7512 \hfill \cr
{Y}_{s}\hfill&0.2470 \hfill &0.2597 \hfill &0.2308 \hfill \cr
{Z}_{s}\hfill&0.01798 \hfill &0.01574 \hfill
&0.0170 \hfill \cr
\overline{N}_s{(^{12}C})\hfill&8.55\hfill &8.50 \hfill&8.55 \hfill \cr
\overline{N}_s{(^{14}N})\hfill&7.97 \hfill &7.92 \hfill&7.97\hfill \cr
\overline{N}_s{(^{16}O})\hfill&8.87 \hfill &8.82 \hfill&8.87 \hfill \cr
\overline{N}_s{(^{20}Ne})\hfill&8.08 \hfill &8.03 \hfill&8.08 \hfill
\cr
\overline{X(\geq^{24}Mg)/{Z}_s}\hfill&  &0.00414 \hfill&
0.00415 \hfill \cr
{R}_{conv}~[R/R_{\odot}]\hfill&0.712 \hfill &0.7105 \hfill &0.7301
\hfill \cr
{T}_{bc}~[{10}^{6}{\rm K}]\hfill&2.20 \hfill &2.10 \hfill &2.105
\hfill \cr
{T}_{eff}~[{\rm K}]\hfill& \hfill &5920 \hfill &5803
\hfill \cr}$$
 
\vfill
\eject
 
{\bf Table V:} Comparison between the solar neutrino fluxes
calculated from the best standard solar model with no diffusion of
BP95
and those calculated with the Dar-Shaviv SSM code with the same
nuclear reaction rates, opacities, composition and astrophysical
parameters. The predictions of the current best standard solar
models of
Dar \& Shaviv and of Bahcall \& Pinsonneault are also included.
 
$$\matrix{ & {\rm BP95(ND)}& {\rm  DS(BPND)} & {\rm BP95}&
{\rm
DS95}\cr
{\phi }_{\nu }(pp)~[{10}^{10}{cm}^{-2}{s}^{-1}]\hfill
&6.01\hfill &6.08\hfill &5.91\hfill &6.10\cr
{\phi }_{\nu }(pep)~[{10}^{8}{cm}^{-2}{s}^{-1}]\hfill
&1.44\hfill &1.43\hfill &1.40\hfill &1.43\cr
{\phi }_{\nu }(^{7}{Be})~[{10}^{9}{cm}^{-2}{s}^{-1}]\hfill &
4.53\hfill &4.79\hfill &5.15\hfill &3.71\cr
{\phi }_{\nu }(^{8}{B})~[{10}^{6}{cm}^{-2}{s}^{-1}]\hfill &
4.85\hfill &5.07\hfill &6.62\hfill &2.49\cr
{\phi }_{\nu }(^{13}{N})~[{10}^{8}{cm}^{-2}{s}^{-1}]\hfill &
4.07\hfill &4.05\hfill &6.18\hfill &3.82\cr
{\phi }_{\nu }(^{15}{O})~[{10}^{8}{cm}^{-2}{s}^{-1}]\hfill &
3.45\hfill &3.38\hfill &5.45\hfill &3.74\cr
{\phi }_{\nu }(^{17}{F})~[{10}^{6}{cm}^{-2}{s}^{-1}]\hfill &
4.02\hfill &4.06\hfill &6.48\hfill &4.53\cr
\Sigma (\phi \sigma)_{Cl}~[SNU]\hfill
&\rm 7\pm 1\hfill &\rm 7\pm 1 \hfill&\rm 9.3\pm 1.4\hfill&
\rm 4.1\pm 1.2\hfill \cr
\Sigma(\phi\sigma)_{Ga}~[SNU]\hfill &\rm 127\pm 6\hfill
&\rm 128\pm 7\hfill &\rm 137\pm 8\hfill
&\rm 115\pm 6\hfill \cr}$$
 
\vfill
\eject
 
\end{document}